\documentstyle[sprocl,epsf]{article}

\newcommand{\piccie}[2]{\setlength{\epsfysize}{#2}\epsffile{#1}}

\def\be{\begin{equation}}
\def\ee{\end{equation}}
\def\bea{\begin{eqnarray}}
\def\eea{\end{eqnarray}}

\newcommand{\bm}[1]{\mbox{\boldmath $#1$}}
\newcommand{\bsigma}{\mbox{\boldmath $\sigma$}}
\newcommand{\bomega}{\mbox{\boldmath $\Omega$}}
\newcommand{\s}{\varsigma}
\newcommand{\minus}{\!-\!}

\begin{document}
 
\title{Spin Glasses}

\author{David Sherrington }

\address{Department of Physics, University of Oxford, Theoretical Physics, 1 Keble Road, Oxford, OX1 3NP, UK\\E-mail: D.Sherrington1@physics.ox.ac.uk}




\maketitle
\abstracts{An introduction and overview is given of the theory of spin glasses and its application.}

\section{Introduction}

The expression ``spin glass'' was originally coined to describe some magnetic alloys in which there was observed non-periodic ``freezing'' of the orientations of the magnetic moments (or ``spins''), coupled with slow response and linear low-temperature heat capacity characteristic of conventional glasses.  The attempt to understand the cooperative physics of such alloys has exposed many previously unknown and unanticipated fundamental concepts and led to the devising of new analytical, experimental and computer simulational techniques. These have had major ramifications throughout the whole field of study of problems involving assemblies of strongly interacting individual entities in which competitive forces yield complex cooperative behaviour.  Such problems are ubiquitous, not only throughout the breadth of condensed-matter physics but also biology, evolution, organizational dynamics, hard-optimization, and environmental and social structures.  In consequence the expression ``spin glass'' has now taken on a wider interpretation to refer to complex glassy behaviour arising from a combination of quenched disorder and competitive interactions or constraints, and to systems exhibiting such behaviour.

In these lectures I shall introduce and partially overview the key concepts and main theoretical techniques for understanding and quantifying model spin glasses, with only limited discussion of experiments and of history and no attempt at completeness.

Despite the caveat of the last sentence, a brief historical introduction to set the scene does seem appropriate \cite{hist}.  The earliest experiments drawing attention to spin glasses as potentially interesting systems were performed on substitutional alloys of magnetic and non magnetic metals.  Evidence for random freezing was provided by the combination of the observation beneath a characteristic temperature of M\"ossbauer line-splitting in zero applied field, indicating a local hyperfine field due to local freezing of the magnetic moments, and the absence of any corresponding magnetic Bragg peak in neutron diffraction, demonstrating that the freezing is not periodic.  In fact, even earlier, measurements of the susceptibility had shown a maximum at a similar temperature, again indicative of non-ferromagnetic freezing, but the observed peak was rounded, suggestive of sluggish response.  The possibility of a true phase transition was highlighted  later when a.c. susceptibility measurements in which static fields were kept very low showed sharp cusps \cite{CM}.  The attempt by Edwards and Anderson \cite{EA} to produce a theory of this transition was the spark for a theoretical revolution which still continues.

Random freezing is certainly a novelty and, as noted above, it was the attempt to explain the occurrence of this feature which stimulated discoveries of great subtlety.  However, a second set of early experiments exposed the presence of another, even more remarkable feature.  These experiments demonstrated severe preparation-dependence effects and considerable slowing-down of response to external perturbations beneath the same characteristic temperature as that found in the susceptibility experiments.  Although its significance was not appreciated until later, one of these striking preparation-dependent features was observed in a d.c. susceptibility experiment; the susceptibility obtained by cooling the system in the measurement field (FC: field cooling) yielded a higher value than that obtained by first cooling in zero field and then applying the measurement field (ZFC: zero field cooling) \cite{TT}.  Figure 1 shows a more recent demonstration \cite{NKH}.   Similarly dramatic preparation-dependence was observed in measurements of the remanent magnetization, which, if measured by cooling in the field and then removing it (TMR: thermoremanent magnetization), is greater than the isothermal remanent magnetization (IRM) obtained by first cooling, then applying and finally removing the field (see fig. 2).  These observations demonstrate that in the new ``phase'' there are many metastable states whose relative free energies vary in different ways with external perturbations and which have significant (free) energy barriers impeding motion from one state to another.

\begin{figure}[tb!]
\centerline{\piccie{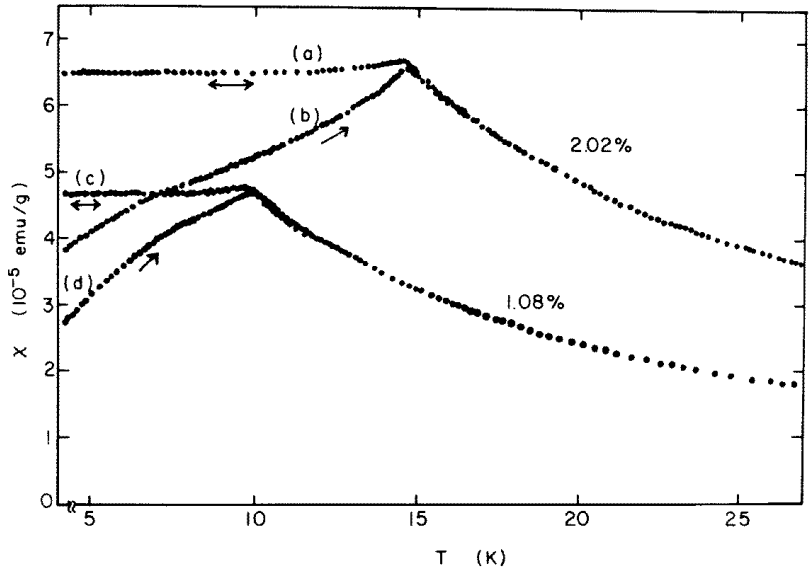}{3in}}
\caption{D.c. susceptibility measurements$^5$ for two \underline{Cu}Mn alloys with 1.08 and 2.02 at \% Mn. Curves (a) and (c) were obtained by cooling in the measurement field (FC),(b) and (d) are the results of zero-field-cooled (ZFC) experiments; from Nagata et. al.$^5$}
\end{figure}

\begin{figure}[tb!]
\centerline{\piccie{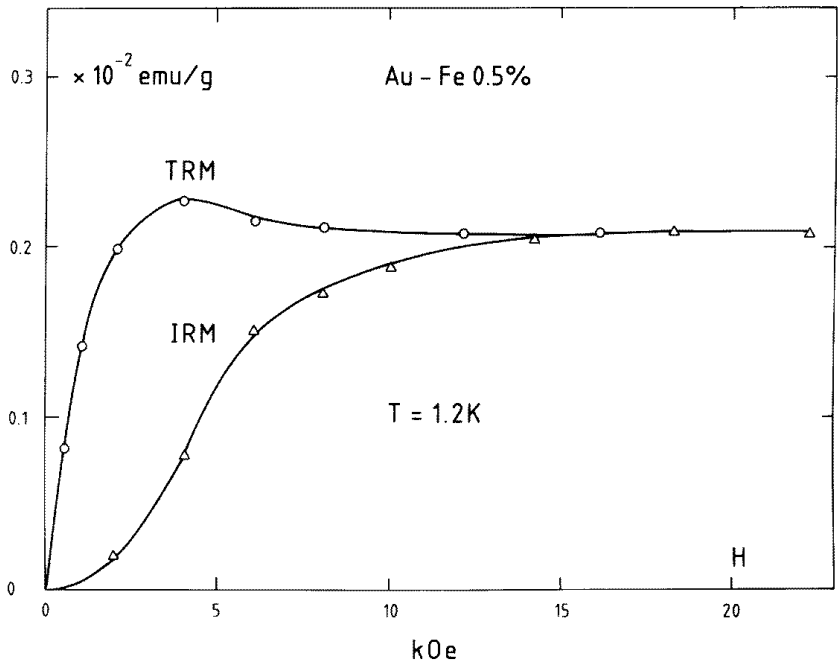}{3.5in}}
\caption{Remanent magnatization as measured in Au-0.5\% Fe at 1.2 K. IRM denotes isothermal remanent magnetization, TRM denotes thermoremanent magnetization; from Tholence and Tournier$^4$}
\end{figure}

\section{The key ingredients}

Spin glass behaviour seems to require two essential ingredients.  These are quenched disorder and frustration \footnote{In fact, recent studies have shown that the disorder can be effectively self-generated and are leading to new insight into conventional glasses, but for the moment we shall ignore this subtlety.}

``Quenched disorder'' refers to constrained disorder in the interactions between the spins and/or their locations.  The spin orientations themselves are variables, (i.e. not constrained), governed by the interactions, external fields and thermal fluctuations, free to order or not as their dynamics or thermodynamics tells them.  The spin glass phase is an example of spontaneous cooperative freezing (or order) of the spin orientations in the presence of the constrained disorder of the interactions or spin locations.  It is thus ``order in the presence of disorder''.   At a deeper level, in real solids the time scale for the ordering of the spin orientations is short but that of ordering of the atoms or interactions is very long.

``Frustration'' refers to conflicts between interactions, or other spin-ordering forces, such that not all can be obeyed simultaneously.

These features are readily visualized in a simple model Hamiltonian appropriate to an idealization of magnetic interactions between atoms with well-defined local moments.
\begin{equation}
H=-\sum_{(ij)}J_{ij}{\bm{S}}_i{\bm{.}}{\bm{S}}_j,
\end{equation}
where the $i,j$ label the magnetic moments/spins, the ${\bm{S}}_i$ are the corresponding spin orientation vectors, and $J_{ij}$ measures the ``exchange'' interaction between the pair of spins $(ij)$.  The variables are the $\{{\bm{S}}_i\}$, while the $\{J_{ij}\}$ are quenched/constrained.  In conventional experimental spin glass systems, the spin locations ${\bm{R}}_i$ are randomly located (on a lattice in the case of the substitutional alloys mentioned earlier) and $J_{ij}$ is a function of $({\bm{R}}_i - {\bm{R}}_j)$ which oscillates in sign with separation.  Frustration arises in that pairs of spins get different ordering instructions through the various paths which link $i$ and $j$, either directly or via intermediate spins.  For theoretical studies it is usually more convenient to consider models in which there are magnetic spins on all sites but the interactions $J_{ij}$ are randomly positive or negative (and quenched); that such models still capture the essence is borne out by computer simulations which yield results qualitatively similar to those of real experiments.  

A natural corollary of this recognition of the key ingredients is that experimentally the behaviour should not be restricted to metallic systems, provided one has disorder and frustration, and indeed the effects have now been seen in several insulating alloys.  The canonical insulating example is $Eu_xSr_{1-x}S,$ in which only $Eu$ is magnetic and for which the nearest-neighbour interactions are ferromagnetic, next-nearest interactions are antiferromagnetic.

The relevance of frustration is that it leads to degeneracy or multiplicity of compromises; for example if one has a set of Ising spins at the corners of a polygon with nearest-neighbour interactions randomly $\pm J$, then one cannot satisfy all the bonds if an odd number of them are antiferromagnetic and the ground state is degenerate.  In extensive systems frustration can have major consequences on cooperative ordering.  For example, a triangular antiferromagnet remains paramagnetic at all temperatures.  Other periodically frustrated systems do order, in a rich plethora of compromise phases, both commensurate and incommensurate with the underlying lattice, but with sufficient quenched disorder no periodic compromise is possible.

\section{Analytic theory: thermodynamics}

Let us now turn to an analysis of spin glasses \cite{TR}, taking as prototype the random-bond Ising model with Hamiltonian
\begin{equation}
H=-\sum_{ij}J_{ij}\sigma_i\sigma_j; ~~\sigma_i=\pm 1.\label{eq:sgH}
\end{equation}
where the $i$ label the spins and the $J_{ij}$ are drawn randomly from distributions $P(J_{ij})$ which are the same for all equivalent pairs of spin locations.  Initially, we shall consider the spins to be on a lattice of sites $(i,j)$.

The possibility of random ordering without ferromagnetism can be hinted at by a straightforward extension of conventional mean-field theory.  Allowing for the lack of spatial symmetry, but ignoring self/cavity field and thermodynamic fluctuation effects, a simple extension of a conventional approximation yields the set of self-consistent mean-field equations
\begin{equation}
\langle\sigma_i\rangle = \tanh(\sum_jJ_{ij}\langle\sigma_j\rangle/kT),
\end{equation}
where the $\langle~\rangle$ brackets indicate a (possibly symmetry-broken) thermodynamic average
\begin{equation}
\langle O\rangle = \frac{TrO \exp(-\beta H)}{Tr \exp(-\beta H)}; ~ ~ Tr\equiv \sum_{\{\sigma_{i}\}=\pm1}; ~ ~ \beta =(kT)^{-1}.
\end{equation}

Conventionally, within mean-field approximation, the critical temperature $T_c$ for the onset of frozen order is given by the existence of a non-trivial solution $(\langle\sigma\rangle\not=0)$ to the linearized mean-field equation.  In the present case this corresponds to
\begin{equation}
\langle\sigma_i\rangle = \sum_j J_{ij}\langle\sigma_j\rangle/kT_c.\label{eq:mf1}
\end{equation}

Averaging over sites and bonds and (without justification) ignoring correlations between those averages yields
\begin{equation}
[\langle\sigma_i\rangle]=\sum_j[J_{ij}][\langle\sigma_j\rangle]/kT_c,\label{eq:mf2}
\end{equation}
where the [~] brackets denote an average over the bond or site disorder. For a symmetric exchange distribution $[J_{ij}]$ is zero and there is no non-trivial solution to (\ref{eq:mf2}) at finite $T_c$.  There is no ferromagnetic solution.  An analogous consideration similarly eliminates any other periodic order in this case.  If, however, (\ref{eq:mf1}) is first squared on each side and then averaged, again ignoring correlations and using $[J_{ij}]=0$, there results
\begin{equation}
[\langle\sigma_i\rangle^2]=\sum_i[J^2_{ij}][\langle\sigma_j\rangle^2]/(kT_c)^2,
\end{equation}
which has a non-trivial solution at a critical temperature given by $kT_c=(\sum_jJ^2_{ij})^{1/2}$ with `order parameter'' $[\langle\sigma_i\rangle^2]$.  Thus, beneath this temperature one has a ``frozen-spin'' state but without periodic order.  Of course, even within mean-field theory this analysis is inadequate in detail, but, as we shall see below, sophisticated mean-field theory also yields a frozen-spin state without periodic order.

\subsection{Replica theory}

Much of the further progress in understanding and quantifying the spin glass problem has employed an artifice known as replica theory.  This was introduced by Edwards and Anderson \cite{EA} to aid in the analysis of physical averages over quenched disorder.

We are generally interested in statistically representative quantities, rather than specific instances.  It is therefore of interest to look at averages over specific disorder.  Indeed, it is a traditional tenet that physical macroscopic measurements on nominally equivalent random systems are overwhelmingly dominated by their mean values; for example, the susceptibility per unit mass of one piece of ${\bm{Cu}}Mn$ alloy is expected to be the same as another of the same concentration, provided they are similarly prepared and subject to the same external fields, even though the precise locations of the individual atoms differ.  Recently, we have come to realize that subtleties of macroscopic fluctuations are possible and we shall reconsider this assumption later.  For the present, however, we shall study the disorder-averaged system.  It is, however, important that one averages physical observables if one is to obtain physically relevant results.  Thus it is sensible to average the free energy $F=-kT\ln Z$, where $Z$ is the partition function
\begin{equation}
Z=Tr\exp(-\beta H),
\end{equation}
but not to average $Z$ itself.  Unfortunately, whereas $Z$ is relatively easy to average, being a sum of exponentials, $\ln Z$ is much harder to average.  The replica trick is an artifice to transform the hard average over $\ln Z$ into an easier one over an effective $Z$.

In particular, the replica trick starts with the mathematical identity 
\begin{equation}
\ln Z=\lim_{n\rightarrow 0} \frac{1}{n}(Z^n-1).
\end{equation}
Then $Z^n$ may be interpreted as the partition function of $n$ identical replicas of the original system.  Introducing a replica label $\alpha = 1,...,n,~Z^n$ may be written as 
\begin{equation}
Z^n=Tr_n\exp\left(-\sum^n_{\alpha=1}H^\alpha/kT\right),
\end{equation}
where $H^\alpha$ is the Hamiltonian with dummy variables labelled by an extra index $\alpha$ and $Tr_n$ is the trace over all the variables.  Because $Z^n$ is a sum over exponentials it is relatively straightforward to average in terms of cumulants.  For the random-bond Ising model one obtains
\begin{equation}
[Z^n]=Tr_n\exp\left\{\sum_{(ij)}\sum_r[J^r_{ij}]_c/(kT)^r \sum_{\alpha,\beta,...}(\sigma^\alpha_i\sigma^\beta_i\cdots\sigma^{\delta r}_i)(\sigma^\alpha_j\sigma^\beta_j\cdots\sigma^{\delta r}_j)\right\}, 
\end{equation}
where $[J^r_{ij}]_c$ indicates the $r$th cumulant moment of $J_{ij}$ and $\alpha,\beta,...\delta_r$ indicate $r$ replica labels, each taking values from 1 to $n$.

At this point it is convenient to follow Edwards and Anderson \cite{EA}(EA) and Sherrington and Southern \cite{SS}(SS) and restrict $P(J)$ to a Gaussian form, since then all the cumulant moments higher than $r=2$ vanish.  To further restrict the problem to single parameters to characterize each of ferromagnetic and spin glass tendencies, the range $(ij)$ can be restricted either to nearest neighbours (as EA and SS) or to all spins (Sherrington-Kirkpatrick \cite{SK}).  Then 
\begin{equation}
[F]=-kT\lim_{n\to o}\frac{1}{n}\left\{Tr_n\exp\left[\sum_{(ij)}(\beta \tilde{J}_o \sum_\alpha\sigma^\alpha_i\sigma^\alpha_j + \beta^2\tilde{J}^2\sum_{\alpha,\beta}\sigma^\alpha_i\sigma^\beta_i\sigma^\alpha_j\sigma^\beta_j)       \right]  -1                  \right\},\label{eq:fenergy}
\end{equation}
where $\tilde{J}_o$ is the mean and $\tilde{J}$ the standard deviation of the $P(J_{ij})$.  Thus we have replaced the original disordered system of eq. (3.1) by one with a  temperature-dependent effective Hamiltonian
\begin{equation}
H_{eff} = - \sum_{(ij)} \left(\tilde{J}_o\sum^n_{\alpha=1}\sigma^\alpha_i\sigma^\alpha_j+\beta\tilde{J}^2\sum^n_{\alpha,\beta=1}\sigma^\alpha_i\sigma^\beta_i\sigma^\alpha_j\sigma^\beta_j\right)
\end{equation}
with no disorder but involving higher-dimensional spins with more complicated interactions and requiring analysis in the limit $n\to 0$.

\subsection{Replica mean-field theory}

For short-range interactions the expression in (\ref{eq:fenergy}) cannot be evaluated exactly.  By analogy with conventional magnetism, it is natural to consider first a mean-field approximation in which an interacting problem is replaced by an effective non-interacting system with self-consistently determined ``fields''.  For infinite-ranged systems with appropriate scaling of $\tilde{J}_o,\tilde{J}$ with the number of spins $N$ such an analysis can be performed exactly, and in consequence the SK model and its variants have become the defining mean field models.  These infinite-ranged models are usually treated by a procedure of mapping to macroscopic variables with extremally dominated generating functionals, but for orientational purposes I shall first follow a route closer to conventional mean-field theory.  Thus one replaces
\begin{equation}
\sum_{ij}\sigma^\alpha_i\sigma^\alpha_j \to \sum_{ij}(\sigma^\alpha_im^\alpha_j - m^\alpha_im^\alpha_j/2); ~ ~ m^\alpha_i = \langle\sigma^\alpha_i\rangle_n,
\end{equation}
\begin{equation}
\sum_{(ij)}\sigma^\alpha_i\sigma^\beta_i\sigma^\alpha_j\sigma^\beta_j \to \sum_{ij}(\sigma^\alpha_i\sigma^\beta_iq^{(\alpha\beta)}_j - q_i^{(\alpha\beta)}q_j^{(\alpha\beta)}/2); q_i^{(\alpha\beta)} = \langle\sigma^\alpha_i\sigma^\beta_i\rangle_n, \alpha\not=\beta,
\end{equation}
where $\langle~\rangle_n$ refers to a thermodynamic average against the effective Hamiltonian and the order parameters $m^\alpha_i, q^{(\alpha\beta)}_i$ are to be determined self-consistently.  Since eq. (13) is translationally invariant, we can assume a similar property for $m^\alpha_i,q_i^{(\alpha\beta)}$. \footnote{provided $J_o$ is non-negative; if $J_o$ were negative we would need to allow for antiferromagnetic order instead.}

The evaluation of $[F]$ now becomes a single-site problem:
\begin{eqnarray}
[F]=  & - &  NkT \lim_{n\to 0}\frac{1}{N}\{Tr_n\exp[\beta\tilde{J}_oz\sum_\alpha(\sigma^\alpha m^\alpha - (m^\alpha)^2/2) \nonumber \\
 & + &  (\beta\tilde{J})^2 z(n+2\sum_{(\alpha\beta)}(\sigma^\alpha\sigma^\beta q^{(\alpha\beta)} - q^{(\alpha\beta)2}/2)) ] - 1 \},\label{eq:fenergy2}
\end{eqnarray}
where $z$ is the coordination number, the trace is now single-site, and $m^\alpha$ and $q^{(\alpha\beta)}$ are given by 
\begin{equation}
m^\alpha =\frac{Tr_n\sigma^\alpha\exp(-\beta\tilde{H}_n)}{Tr_n\exp(-\beta\tilde{H}_n)}, ~ ~ ~ ~ ~ ~ ~ ~ ~ ~ 
q^{(\alpha\beta)} =\frac{Tr_n\sigma^\alpha\sigma^\beta\exp(-\beta\tilde{H}_n)}{Tr_n\exp(-\beta\tilde{H}_n)}\label{eq:mf3}
\end{equation}
where $-\beta\tilde{H}_n$ is the argument of the exponential in (\ref{eq:fenergy2}), or, equivalently, by the extremal equations
\begin{equation}
\frac{\delta\tilde{F}_n}{\delta m^\alpha} = \frac{\delta\tilde{F}_n}{\delta q^{(\alpha\beta)}} = 0
\end{equation}
where 
\begin{equation}
\tilde{F}_n=-kT\ln Tr_n\exp(-\beta\tilde{H}_n).
\end{equation}

\subsection{Replica-symmetric Ansatz}

Equations (\ref{eq:mf3}) represent a set of self-consistent equations.  Were $n$ a fixed finite integer, they would be straightforward to solve \cite{S} but with the limiting procedure $n\to 0$ the problem is more difficult, requiring an appropriate analytic continuation.  For this reason a further simplifying Ansatz was proposed$^3$, the so-called replica-symmetric (RS) Ansatz in which one assumes 
\begin{equation} m^\alpha=m, ~{\rm{all}}~\alpha, ~ ~ ~ ~ ~ ~ ~ ~ ~ q^{(\alpha\beta)} = q, ~{\rm{all}}~\alpha\not=\beta.
\end{equation}
This Ansatz is natural, since the replicas are mathematical artifices and are apparently indistinguishable.  In fact, however, the situation is more subtle, but we defer further discussion of the subtlety until later.

The computational advantage of the replica-symmetric Ansatz is two-fold; firstly, it permits (\ref{eq:fenergy2}) to be expressed in terms of independent spin terms, which may be evaluated readily, and secondly, it provides for straightforward analytic continuation.  To see the first of these properties, note that the ``interacting'' term in the exponent
\begin{equation}
2\sum_{\alpha\beta}\sigma^\alpha\sigma^\beta q^{(\alpha\beta)} ~ ~\to ~ ~  q\{(\sum_\alpha\sigma^\alpha)^2-1\}
\end{equation}
which can be reduced to single-$\sigma$ form via the identity
\begin{equation}
\exp(\mu x^2) = (2\pi)^{-1/2}\int dy\exp(-y^2/2+(2\mu)^{1/2}xy);\label{eq:squares}
\end{equation}
here we take $\mu=(\beta\tilde{J})^2zq$ and $x=\sum\sigma^\alpha$.  The simple analytic continuation occurs because now all $\sigma^\alpha$ enter linearly with the same coefficient.  Evaluation of the trace and taking of the limit $n\to 0$ yields
\begin{equation}
[F]=N\left\{\frac{Jm^2}{2}-\frac{\beta J^2}{4}(1-q)^2 - kT \int dhP(h)\ln (2\cosh\beta h)\right\},
\end{equation}
where $J_0=\tilde{J}_0z,J=\tilde{J}z^{1/2}$ and $m$ and $q$ are determined self-consistently from 
\begin{equation}
m=\int dh P(h)\tanh \beta h, ~ ~ ~ ~ ~ ~ ~ ~ ~ ~ 
q=\int dh P(h)(\tanh\beta h)^2,\label{eq:mqRS}
\end{equation}
with
\begin{equation}
P(h)=(2\pi Jq^2)^{-1/2}\exp(-(h-J_om)^2/2Jq^2). \label{eq:PRS}
\end{equation}

Within this approximation $m$ and $q$ can also be identified as
\begin{equation}
m=[\langle\sigma_i\rangle], ~ ~ ~ ~ ~  q=[\langle\sigma_i\rangle^2],
\end{equation}
i.e. as the average magnetization and mean-square disorder-averaged local moment \cite{SK} \footnote{This identification (and extensions of it) follows from writing $\langle\sigma_i\rangle^r$ as the trace of $\sigma_i^1 ...\sigma^r_i$ over r replicas of the disordered system $$\langle\sigma_i\rangle^r = \frac{Tr_r\sigma_i^1\cdots\sigma^r_i \exp(-\beta(H^1 + \cdots + H^r))}{Tr_r\exp(-\beta(H^1+\cdots+H^r))},$$ multiplying numerator and demoninator by $Z^{n-r}$, taking the limit $n \to 0$ and averaging over the disorder to yield $$[\langle\sigma_i\rangle^r]=\lim_{n\to 0} \langle\sigma_i^{\alpha_{1}\cdots}\sigma^{\alpha_{r}}_i\rangle_n; \alpha_1\not=\alpha_2\not=\cdots\not=\alpha_r. $$}.  Hence non-zero $q$ implies a cooperatively frozen magnetic state, while non-zero $m$ implies that that frozen state has a ferromagnetic component.  Thus, solutions to eqs. (\ref{eq:mqRS}) and (\ref{eq:PRS}) fall into three possible categories:\vskip0.05in

\noindent(a) paramagnetic if both $m$ and $q$ are zero.\\
(b) ferromagnetic if both $m$ and $q$ are non-zero.\\
(c) spin glass if $q$ is non-zero but $m$ is zero.\vskip0.05in

The resultant phase diagram is shown in fig. 3a, while fig 3b shows for comparison a typical experimental phase diagram.  When account is taken of the different scalings with concentration of effective interactions $J_0,J$ there is a qualitative accord between the figures \cite{SS}.

\begin{figure}[tb!]
\centerline{\piccie{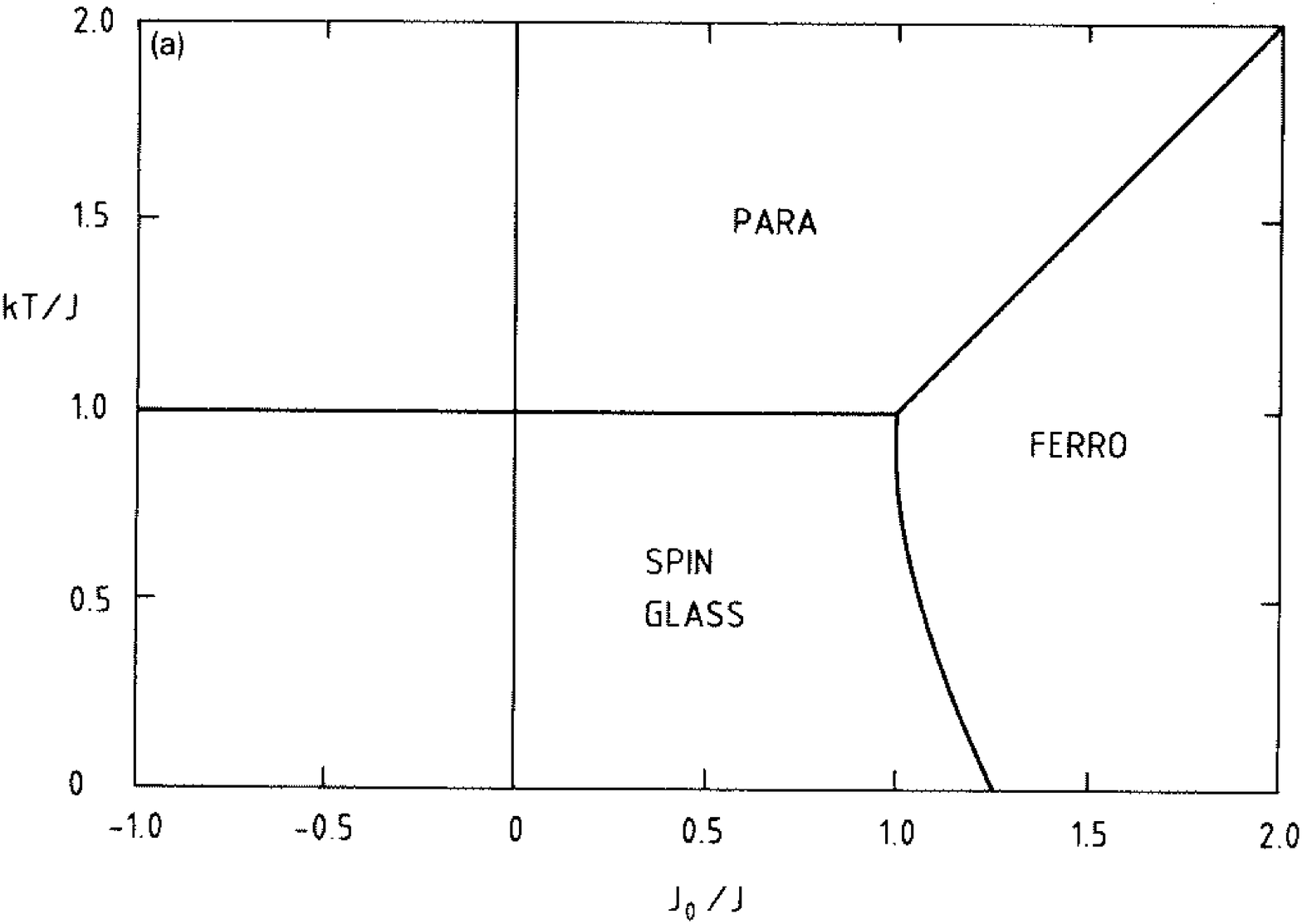}{1.7in}\kern3em\raise3.5pt\hbox{\piccie{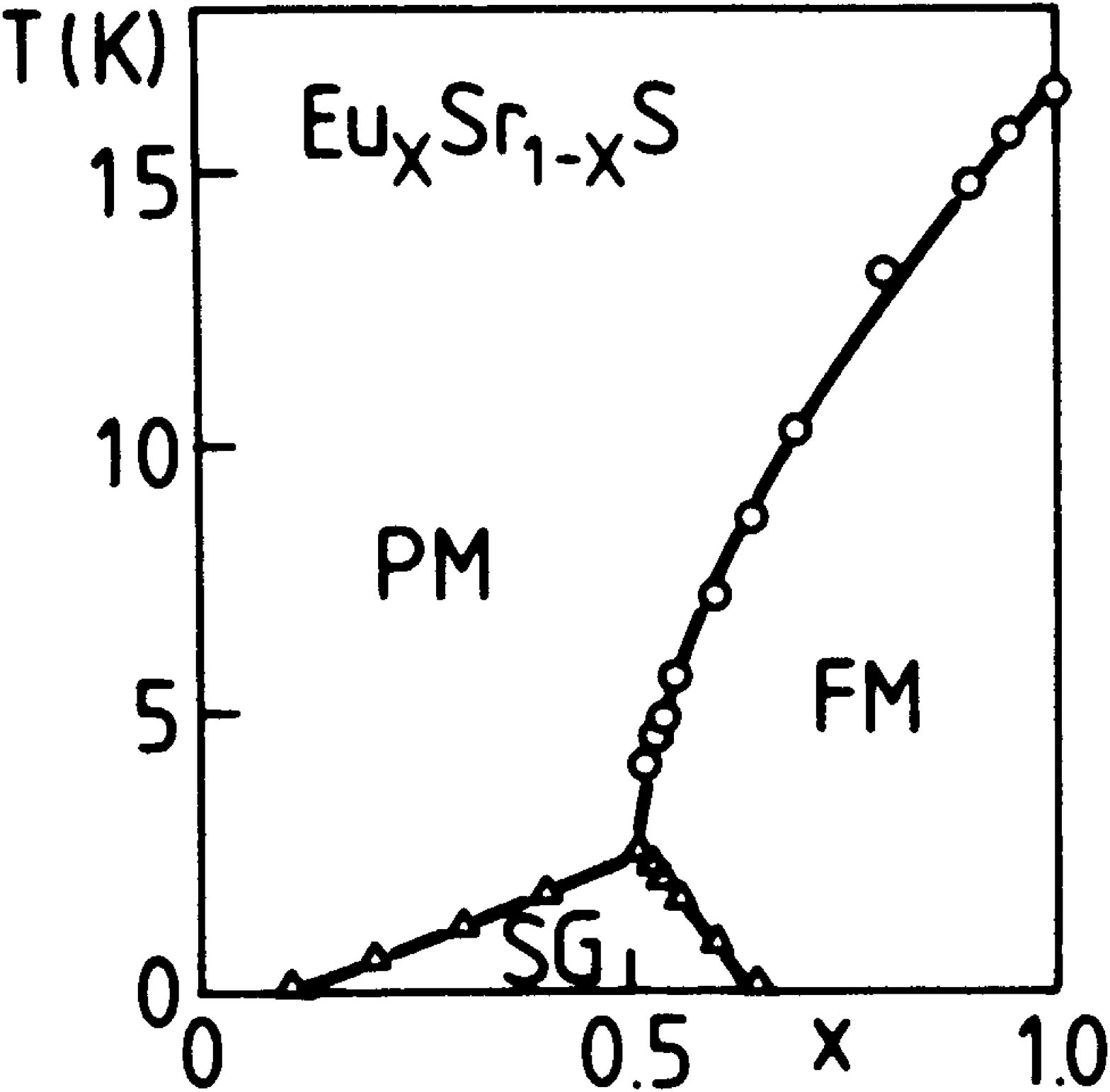}{1.7in}}}
\caption{(a) Phase diagram obtained in replica-symmetric mean-field theory for a random-bond Ising model; (b) Phase diagram of $Eu_xSr_{1-x}S$; from Maletta and Convert 1979$^{13}$.}
\end{figure}

Let us concentrate now on the spin glass phase.  As the temperature is lowered beneath the critical temperature $T_g=J/k,q$ grows continuously from zero as 
\begin{equation}
q=\tau+{1\over 3}\tau^2+0(\tau^3); ~ ~ \tau=(T_g - T)/T_g.
\end{equation}

This yields a cusp in the zero-field susceptibility $\chi(T)$, as is observed experimentally.  This is most readily seen from the fluctuation correlation form for the differential susceptibility:
\begin{equation}
\chi(T) = (kT)^{-1}N^{-1}\sum_{ij}(\langle\sigma_i\sigma_j\rangle - \langle\sigma_i\rangle\langle\sigma_j\rangle).
\end{equation}

For the case $J_0 = 0$, the terms in the summand are zero on average unless $i = j$ and
\begin{equation}
\chi(T)=\chi_0(T) = (NkT)^{-1}\sum_i(1-\langle\sigma_i\rangle^2),\label{eq:susc}
\end{equation}
or passing to the average
\begin{equation}
\chi_0(T) = (kT)^{-1}(1-[\langle\sigma_i\rangle^2]) = (kT)^{-1}(1 - q).
\end{equation}
Hence, in the vicinity of $T=T_g$ one has 
\begin{eqnarray}
\chi_0(T) & = &(kT_g)^{-1}(1-|\tau|+O(\tau^2)), ~ ~ T> T_g,\\
&  & (kT_g)^{-1}(1-{1\over3}\tau^2 + O(\tau^3)), ~ ~ T< T_g.
\end{eqnarray}

For $J_o\not=0, \chi$ has a similar cusp but with an overall enhancement factor giving
\begin{equation}
\chi(T) = \chi_0(T)/(1 - J_0\chi_0(T)).
\end{equation}

In the presence of a finite external field $b, P(h)$ of eq. (\ref{eq:PRS}) is modified by replacing $h\to h-b$ and $\chi(T)$ is rounded, as shown in fig. 4a. A similar rounding of the cusp in $\chi$ is found experimentally (fig. 4b); indeed it is the great sensitivity to such rounding which delayed the experimental discovery of a sharp transition.

\begin{figure}[tb!]
\centerline{\piccie{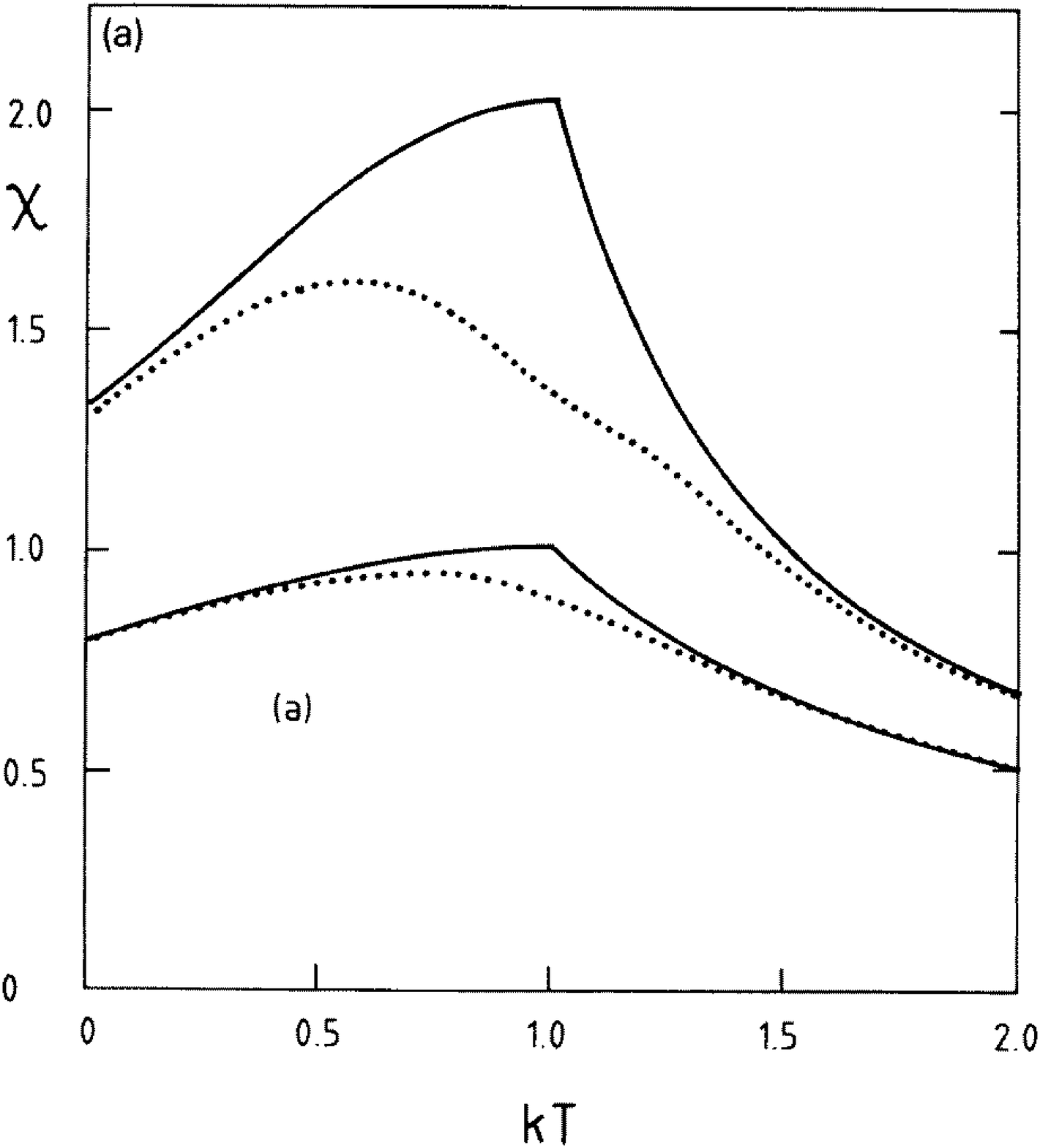}{2in}\kern3em\raise7pt\hbox{\piccie{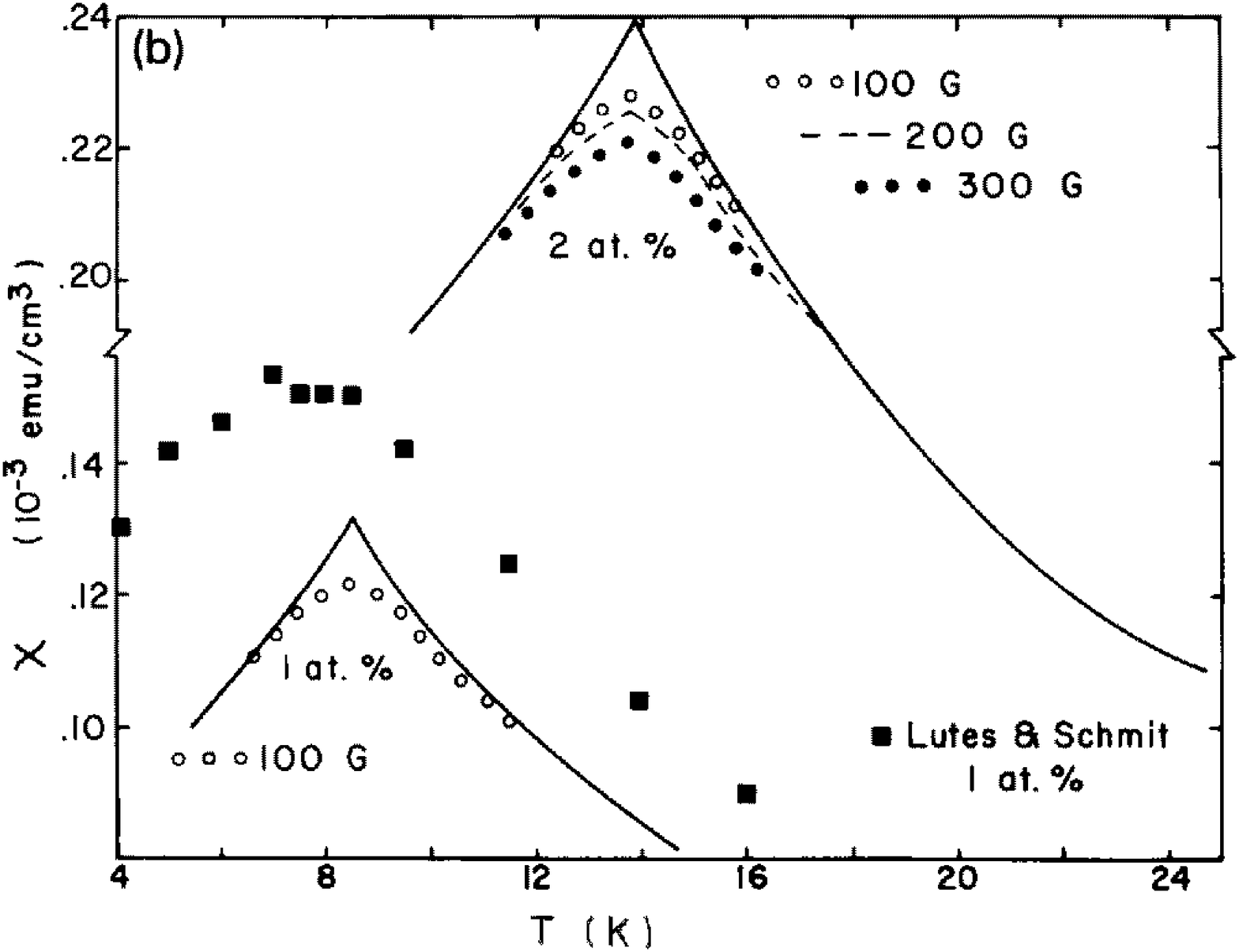}{1.7in}}}
\caption{(a) Differential susceptibility of a random-bond spin glass as given by replica-symmetric mean-field theory.  Solid curves are for zero field, dotted curves for $b=0.1J$.  Curves (a) are for $J_0/J=0$, curves (b) are for $J_o/J=0.5$; Sherrington and Kirkpatrick$^8$. (b) External field dependence of a.c. susceptibility of \underline{Au}Fe; Cannella and Mydosh$^2$.}
\end{figure}

For continuous phase transitions one is used to looking for a diverging response function.  Clearly, the uniform susceptibility is not such a response function for a spin glass transition.  Rather, the divergent susceptibilities are the $q$-response to a random field and the related non-linear uniform susceptibility.  The first of these, sometimes called the spin glass susceptibility, is defined as
\begin{equation}
\chi_{SG}=\partial^2Q/\partial\tilde{b}^2, \label{eq:XSG}
\end{equation}
where
\begin{equation}
Q=\sum\langle\sigma_i\rangle^2_{\tilde{b}} ~ ~ ;
\end{equation}
here $\langle~~\rangle_{\tilde{b}}$ refers to a thermodynamic average against a system with an applied field $\{h_i\}=\{\tilde{b}\xi_i\}$, where the $\xi_i = \pm 1$ are random and quenched.  Formally performing the differentiation of (\ref{eq:XSG}) yields the fluctuation form \cite{FH}
\begin{equation}
\chi_{SG}=\beta^2N^{-1}\sum_{ij}(\langle\sigma_i\sigma_j\rangle - \langle\sigma_i\rangle\langle\sigma_j\rangle)^2.\label{eq:sgsusc}
\end{equation}
For $J_0 = 0$ and for  $T>T_g$ one obtains within mean-field theory 
$\chi_{SG}=(T^2-T^2_g)^{-1}$, which diverges as $T\to T_g$.  For a finite uniform $b$, higher-order susceptibilities can be defined by
\begin{equation}
m=N^{-1}\sum_i\langle\sigma_i\rangle = \chi b + \chi_3b^3 + \chi_5b^5 + \cdots.
\end{equation}
Within the present mean-field theory, the non-linear susceptibilities $\chi_3,\chi_5$ etc. also diverge at $T_g$: in fact, $\chi_3$ and $\chi_{SG}$ are linearly related\cite{Suzuki,Chalupa}.  A similar divergence of these higher-order susceptibilities has been observed experimentally\cite{BMI,MB}, albeit with a critical exponent different from that given by mean-field theory.

In fact, however, the replica-symetric Ansatz is not everywhere stable, even within mean-field theory$^{14}$.  The problem is clearly exposed by looking at the normal modes of fluctuation of $q^{\alpha\beta}$ and $m^\alpha$ in replica space about their replica-symmetric values.  Thus one introduces fluctuation terms $\eta,\epsilon$ by
\begin{equation}
q^{\alpha\beta} = q + \eta^{\alpha\beta}, ~ ~ ~ ~ ~ ~ ~ ~ ~ 
m^\alpha = m + \epsilon^\alpha
\end{equation}
where $q, m$ are the replica-symmetric values, expands the free energy expression $[F]$ to second order in the fluctuations and studies its normal-mode spectrum in the limit $n\to 0$.  Beneath a surface in $(T/J, J_0/J, b/J)$ space, shown in fig. 5 and including all of the spin glass phase, the replica-symmetric Ansatz is found to be unstable against the replica-symmetry breaking $\eta$-modes.  This instability surface is known, after its discoverers, as the de Almeida-Thouless (AT) surface, its projections as AT lines.  

\begin{figure}[tb!]
\centerline{\piccie{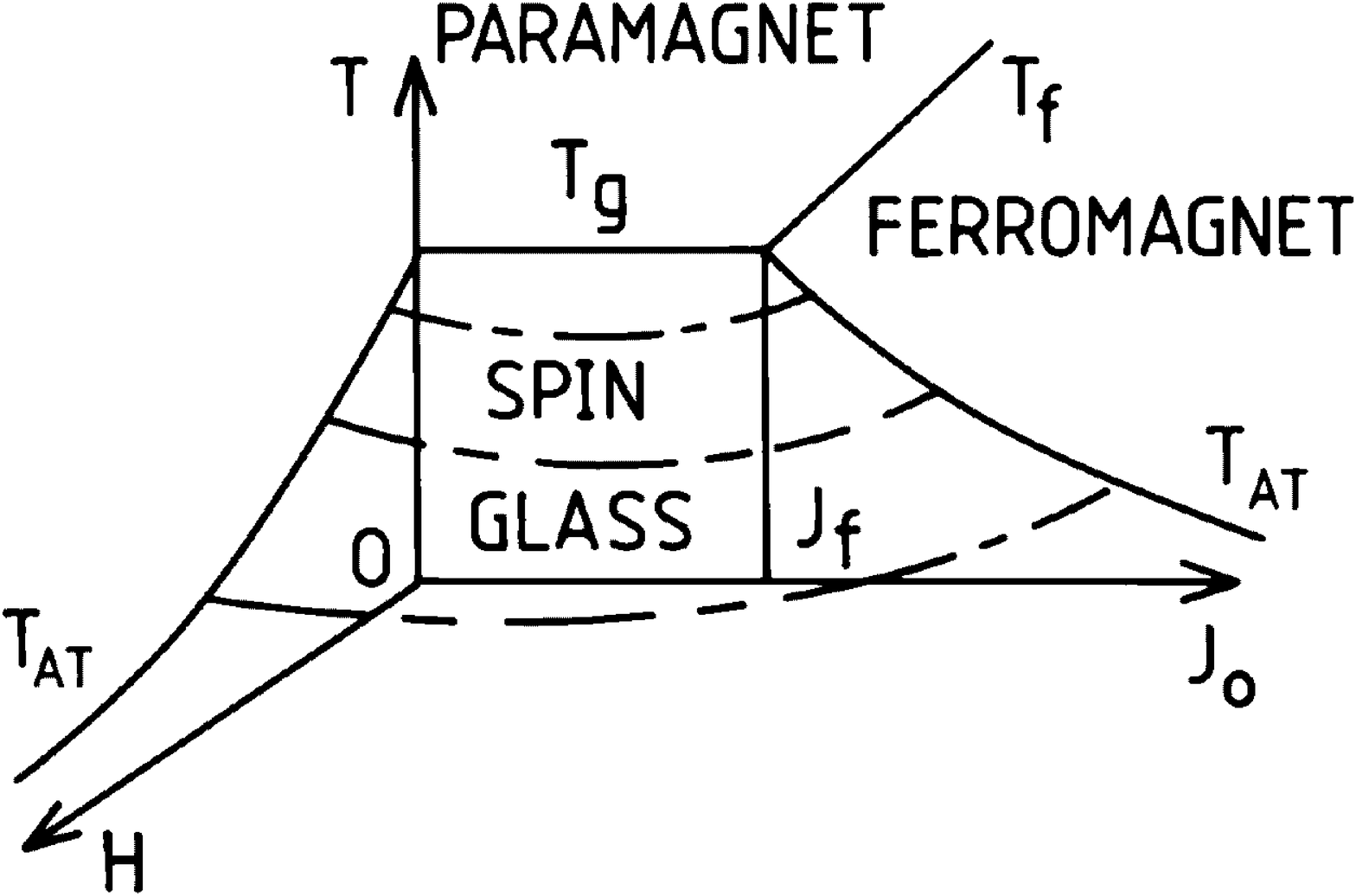}{2in}}
\caption{De Almeida-Thouless surface (indicated by chain-hatching) for the limit of stability of the replica-symmetric Ansatz for mean-field theory for a randon-bond Ising model.  The Ansatz is unstable on the side of the surface closer to the origin.  The phase line beneath the AT surface is calculated within the Parisi Ansatz.}
\end{figure}

The inadequacy of the replica-symmetric Ansatz shows in other ways.  The first of these to be recognized is its prediction of an unphysical (negative) form for the zero-temperature entropy of the random-bond Ising model~\cite{SK}.  Another is the result of an evaluation of the spin glass susceptibility $\chi_{SG}$ beneath $T_g$, taking account of fluctuations to harmonic order, which yields a negative value \cite{PR,BM}, despite the manifestly positive-definite form of (\ref{eq:sgsusc}).

Beneath the AT surface a more subtle Ansatz is required for a stable solution to the mean-field equations (\ref{eq:fenergy2}),(\ref{eq:mf3}).  Various schemes were proposed, whose analysis suggested that many levels of symmetry breaking were necessary to achieve stability.  One due to Parisi \cite{P} has proven to satisfy all stability tests to date.

\subsection{Replica symmetry breaking: Parisi's Ansatz}

In general, the Parisi scheme involved an infinite hierarchy of replica subdivisions \footnote{In fact, some models of current interest have only a single level of sub-division, but the model of Edwards-Anderson/Sherrington-Kirkpatrick requires the full hierarchical structure.}.  It may be formulated as follows.  Consider $q^{\alpha\beta}$ as a symmetric $n \times n$ matrix with zeros on its diagonals and taking the value $q^{\alpha\beta}=q^{(\alpha\beta)}$ on the off-diagonals \footnote{Recall that the notation $(\alpha\beta)$ refers to {\it pairs} of {\it different} labels $\alpha\beta$.}.  The replica-symmetric Ansatz corresponds to taking all the off-diagonal elements to be identical, equal to $q$.  Parisi's method may be viewed as a series of subdivisions.  First the $(n \times n)$ matrix is subdivided into $(m_1 \times m_1)$ blocks and all the elements of the off-diagonal blocks are given the value of $q_1$.  The $(m_1 \times m_1)$ blocks on the diagonal of the $(n \times n)$ matrix are subdivided further into $(m_2 \times m_2)$ sub-blocks and all the elements of the off-diagonal subblocks given the value of $q_2$.  The $(m_2 \times m_2)$ sub-blocks on the diagonal are subdivided further, into $(m_3 \times m_3)$ sub-sub-blocks.  The elements of the off-diagonal $(m_3 \times m_3)$ sub-sub-blocks are given the value $q_3$, while the diagonal sub-sub-blocks are further divided, and so on, giving an infinite regression of subdivisions of the diagonal blocks, with 
\begin{equation}
n \geq m_i \geq m_2 \geq \cdots \geq 1.\label{eq:mseries}
\end{equation}

So far in this discussion of subdivision, we have been envisaging $n,m_i$ as integers, but the next step in the analysis is to consider analytic continuation to $n \to 0$, and with it the inversion of the order of (\ref{eq:mseries}),
\begin{equation}
0 \leq m_1 \leq m_2 \leq \cdots \leq 1,
\end{equation}
together with relabellings with
\begin{equation}
m_k/m_{k+1}\to 1 - dx/x, ~ ~ ~ ~ ~ ~ ~ ~ 
q_k \to q(x), ~ ~  0 \leq x \leq 1.
\end{equation}

Finally, $q(x)$ is treated as a variational-parameter function and the extremum of $[F]$ with respect to $q(x)$ is taken.

Within the region of parameters for which replica symmetry is stable, $q(x)$ is a constant for all $x$, but in the region for which the RS-Ansatz is unstable $q(x)$ is not flat.  For small reduced temperature $\tau = (T_g - T)/T_g$ the leading behaviour of $q(x)$ is shown in fig. 6.

\begin{figure}[tb!]
\centerline{\piccie{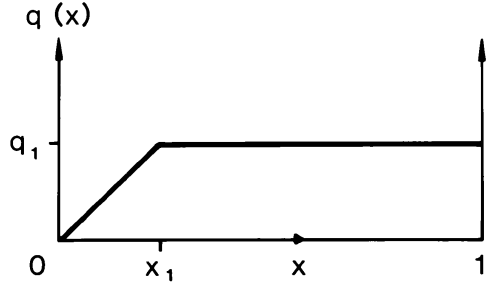}{1.5in}}
\caption{Parisi order function $q(x)$ for the random-bond Ising model in zero external field at small reduced temperature $\tau$.  To leading order in $\tau,q_1=x_1=\tau$}
\end{figure}

Before turning to a discussion of the implications of the Parisi Ansatz for observables such as the susceptibility, let us consider its interpretation \cite{P2}.  To this end, it is useful first to introduce another concept, that of overlap, and of its distribution.

The overlap between two microscopic Ising states $s,s^\prime$ is defined by
\begin{equation}
q^{ss^{\prime}}=N^{-1}\sum_i\sigma^s_i\sigma^{s^{\prime}}_i,
\end{equation}
where $\{\sigma^s_i\}$ is the spin configuration in microstate $s$.  The distribution of {\it microstate} overlaps is given by
\begin{equation}
\tilde{P}(q)=\sum_{s,s^{\prime}} p_sp_{s^{\prime}} \delta(q-q^{ss^{\prime}}),
\end{equation}
where $p_s$ is the probability of the microstate $s$.  In a conventional thermodynamic ensemble
\begin{equation}
p_s = \frac{\exp(-H_s/kT)}{Tr\exp(-H_s/kT)},
\end{equation}
where $H_s$ is the value of the Hamiltonian in state $s$.

The application of the replica procedure to the evaluation of $[\tilde{P}(q)]$, together with mean-field theory using the Parisi Ansatz, yields \cite{P2}
\begin{equation}
[\tilde{P}(q)] = \int^1_0 \delta(q - q(x)) dx \equiv dx/dq.
\end{equation}
Equation (3.53) thus provides an interpretation of $q(x)$ as the inverse of the average microstate distribution.

Similarly one may define the distribution of overlaps between thermodynamic {\it macrostates}.  The overlap distribution between two thermodynamic macrostates $S, S^\prime$ is defined by
\begin{equation}
q^{SS^{\prime}} = N^{-1}\sum_i m^S_i m_i^{S^{\prime}},\label{eq:overlap}
\end{equation}
where $\{m_i^S\}$ is the thermodynamic average of $\{\sigma_i\}$ in macrostate $S$.  The macrostate overlap distribution is given by
\begin{equation}
P(q) = \sum_{SS^{\prime}} P_SP_{S^{\prime}}\delta(q - q^{SS^{\prime}}),
\end{equation}
where $P_S$ is the probability of macrostate $S$,
\begin{equation}
P_S = \frac{\exp(-F_S/kT)}{\sum_{S^{\prime}}\exp(-F_{S^{\prime}}/kT)}
\end{equation}
with $F_S$ the free energy of state $S$.

An important conceptual link arises from the fact the $P(q)$ can be shown to be identical to $\tilde{P}(q)$.  Hence, 
\begin{equation}
[P(q)] = dx/dq
\end{equation}
and $(dx/dq)$ is interpreted as giving the average overlap distribution of thermodynamic macrostates in a Boltzmann-Gibbs ensemble.

A conventional Ising ferromagnet has only two thermodynamic states, spin up and spin down, and
\begin{equation}
P(q) = {1\over2}(\delta(q - m^2) + \delta(q + m^2)),
\end{equation}
where $m$ is the magnetization per spin.  An infinitesimal field suffices to eliminate the peak at $q = - m^2$.  Correspondingly, $q(x)$ is a constant at $q = m^2$ for all $x$.

Similarly, in a region of replica symmetry, in which $q(x)$ is constant as a function of $x$, say with $q(x) = q_0,[P(q)]$ has a single delta function peak, at $q = q_0$, which we interpret as implying a single thermodynamic state.  By contrast, in the region of replica-symmetry breaking, in which $q(x)$ has structure, $[P(q)]$ has weight over a range of $q$, indicating the existence of many thermodynamic states.  For example, for $T$ just smaller than $T_g$ and in the absence of a magnetic field, where $q(x)$ has the form shown in fig. 6, $[P(q)]$ has a delta function at $q = q(1)$, arising from the plateau in $q(x)$, together with a continuum between $q = 0$ and $q = q(1)$, arising from the ramp in $q(x)$.  The delta function is interpreted as the overlap of thermodynamic states with themselves $(S=S^\prime$ in (\ref{eq:overlap})), while the continuum corresponds to a range of overlaps between different non-equivalent macrostates.

This identification of $P(q)$ as an overlap distribution of thermodynamic states leads to the expression of thermodynamic observables as integrals over $q(x)$.  For example, we recall (\ref{eq:susc}) for the zero-field susceptibility of a system with symmetric $P(J_{ij})$:
\begin{equation}
\chi(t) = (kT)^{-1}(1-[\langle\sigma_i\rangle^2]).
\end{equation}
Averages such as $[\langle\sigma_i\rangle^2]$ follow directly from $[P(q)]$, from relations such as 
\begin{equation}
[\langle\sigma_1\sigma_2\cdots\sigma_k\rangle^2] = \int dq~q^k[P(q)] = \int^1_0 dx q(x)^k.
\end{equation}
Hence, in particular, 
\begin{equation}
[\langle\sigma_1\rangle^2]=\int dq~q[P(q)] ~ ~ = \int^1_0 dxq(x),
\end{equation}
so that
\begin{equation}
\chi(T) = (kT)^{-1}\left(1 - \int^1_0 q(x)dx\right).\label{eq:susc2}
\end{equation}

For $T>T_g,q(x)$ is zero and (\ref{eq:susc2}) reproduces the Curie law.  For $T<T_g$, Parisi theory gives
\begin{equation}
1 - \int^1_0 q(x)dx = T/T_g
\end{equation} 
and
\begin{equation}
\chi=(kT_g)^{-1}, ~{\rm all}~ T<T_g.
\end{equation}
This result of Curie-like behaviour above $T_g$ and flat $\chi(T)$ beneath $T_g$ will be seen to match the experimental field-cooled results, as shown in fig. 1.  On the other hand if $q(x)$ is replaced by $q(1)$ in (\ref{eq:susc2}), the resultant $\chi(T)$ is similar to the replica-symmetric result, with $\chi(T)$ decreasing monotonically beneath $T_g$.  This is reminiscent of the zero-field cooled or a.c. susceptibility measurements.  Thus one has apparent correspondences between field cooling and the exploration of all thermodynamic states, and between zero-field cooling and trapping in a single thermodynamic state \footnote{More strictly one needs to consider dynamics, which is the subject of section 4}.

It is interesting to note that the onset of replica-symmetry breaking represents an unusual type of phase transition which can lead to discontinuities in observables even in the presence of a finite conjugate field.  For example, in an external field $q(x)$ is non-zero for all $x$, but as long as the temperature is above that of the AT instability it is not $x$-dependent.  As the temperature is reduced through the AT value, $q(x)$ acquires structure and the Gibbs average and single-state (or field-cooled and zero-field-cooled) susceptibilities start to diverge, despite the fact that the discontinuity in the former in the zero-field limit is removed in a finite field.  The evolution of $q(x)$ with temperature in a finite field $H$ is illustrated in fig. 7, while the resultant susceptibility is shown in fig. 8.  In fig. 7 curve (a) corresponds to a value of the reduced temperature $\tau$ which is less than a critical value $\alpha H^{2/3}$ (where $\alpha = [3/(4J^2)]^{1/3})$ and yields

\begin{figure}[tb!]
\centerline{\piccie{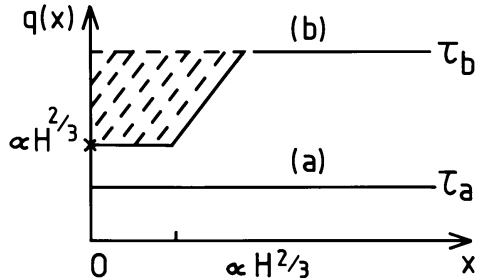}{1.5in}}
\caption{q(x) for a random-bond Ising model in Parisi mean-field theory for (a) $T>T_{AT}$, (b) $T<T_{AT}.$  The hatched area determines the anomaly $\Delta=\chi_{FC} - \chi_{ZFC}$.}
\end{figure}

\begin{figure}[tb!]
\centerline{\piccie{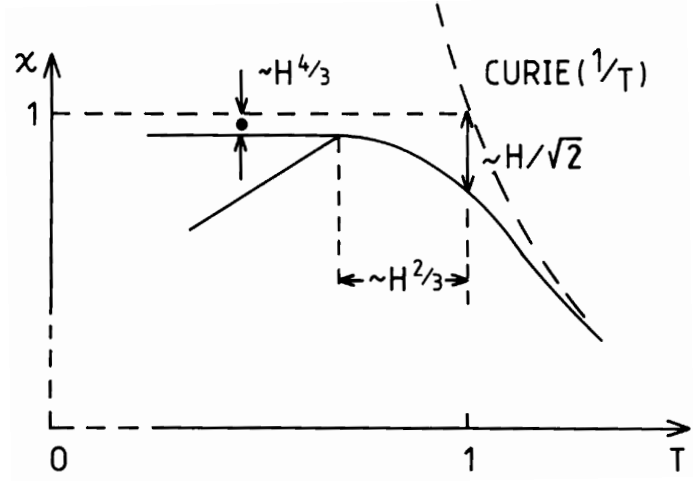}{2.5in}}
\caption{Susceptibility of an Ising spin glass in an applied field $H$, as predicted by Parisi's mean-field theory.  The upper curve shows the full Gibbs average, obtained from the full $q(x)$ and interpreted as the field-cooled (FC) susceptibility.  The lower curve shows the result of restricting to one thermodynamic state, as obtained from $q(1)$ and interpreted as the zero-field-cooled susceptibility.}
\end{figure}
\begin{equation}
q(x) = \tau_a = {1\over2}(\tau + (\tau^2 + 2H)^{1/2} + \cdots).
\end{equation}
As the temperature is lowered $\tau_a$ increases until it reaches $\alpha H^{2/3}$.  Beyond this temperature $q(0)$ is pinned at $\alpha H^{2/3}$ but $q(1)$ continued to rise, as shown in curve (b).  The difference between the two susceptibilities, Gibbs (or FC) and single-state (or ZFC), is given by
\begin{equation}
\Delta=(kT)^{-1}\left(q(1) - \int^1_0 q(x)dx\right)
\end{equation}
and is indicated (without the $(kT)^{-1}$ factor) by the hatched region of Fig. 7.  The susceptibility curves of fig. 8 follow directly from \footnote{For convenience we use the labels FC, ZFC, although strictly the quantities given are the Gibbs ensemble and single-state susceptibilities.}
\begin{equation}
\chi_{FC}=(kT)^{-1}\left(1-\int^1_0 q(x)dx\right),
\end{equation}
and
\begin{equation}
\chi_{ZFC}=(kT)^{-1}(1 - q(1)).
\end{equation}

The Parisi solution exhibits several other interesting features.  In particular we shall mention ultrametricity \cite{RTV,MPSTV} and non-self-averaging \cite{BM2}.\vskip0.10in

\subsection{Ultrametricity}  
A space is described as ``ultrametric'' if distances in that space obey the following property: given three points a, b, c separated by distances $d_{ab}, d_{bc}, d_{ca}$ and labelled such that
\begin{equation}
d_{ab} \geq d_{bc} \geq d_{ca},
\end{equation}
then
\begin{equation}
d_{ab} = d_{bc};
\end{equation}
i.e. the two largest distances between three points are equal.  To relate this to the present study, we note that distance and overlap are complementary; distance is a measure of separation, overlap of similarity.  Thus a space of overlaps is ultrametric if for three states $S, S^\prime, S^{\prime\prime}$ labelled so that
\begin{equation}
q^{SS^{\prime}} \leq q^{S^{\prime}S^{\prime\prime}} \leq q^{S^{\prime\prime}S},
\end{equation}
one has
\begin{equation}
q^{SS^{\prime}} = q^{S^{\prime}S^{\prime\prime}};
\end{equation}
i.e. the two smallest overlaps are equal.

With the definition of overlap given in (\ref{eq:overlap}), Parisi theory predicts such ultrametricity.  A particular interest in this result lies in its implication of hierarchical order.  This implication is obvious if one considers a hierarchical tree as shown in fig. 9, takes its endpoints as the states and their pairwise overlap as determined by how far back in the evolutionary tree one needs to go to find a common ancestor, the overlap being smaller the farther back one needs to go.  One readily sees that for any three states the two smallest overlaps are equal.  This observation of ultrametricity underpins \footnote{This picture was originally conceived as a result of more philosophical considerations and via computer simulations e.g. Kirkpatrick and Sherrington \cite{KS}} the conception of spin glasses as having free-energy surfaces with ``hills-and-valleys'' structure, since if one considers valleys in a mountain landscape as ``states'' and the height of the lowest col between two valleys as their ``separation'', then those separations are ultrametrically ordered.

\begin{figure}[tb!]
\centerline{\piccie{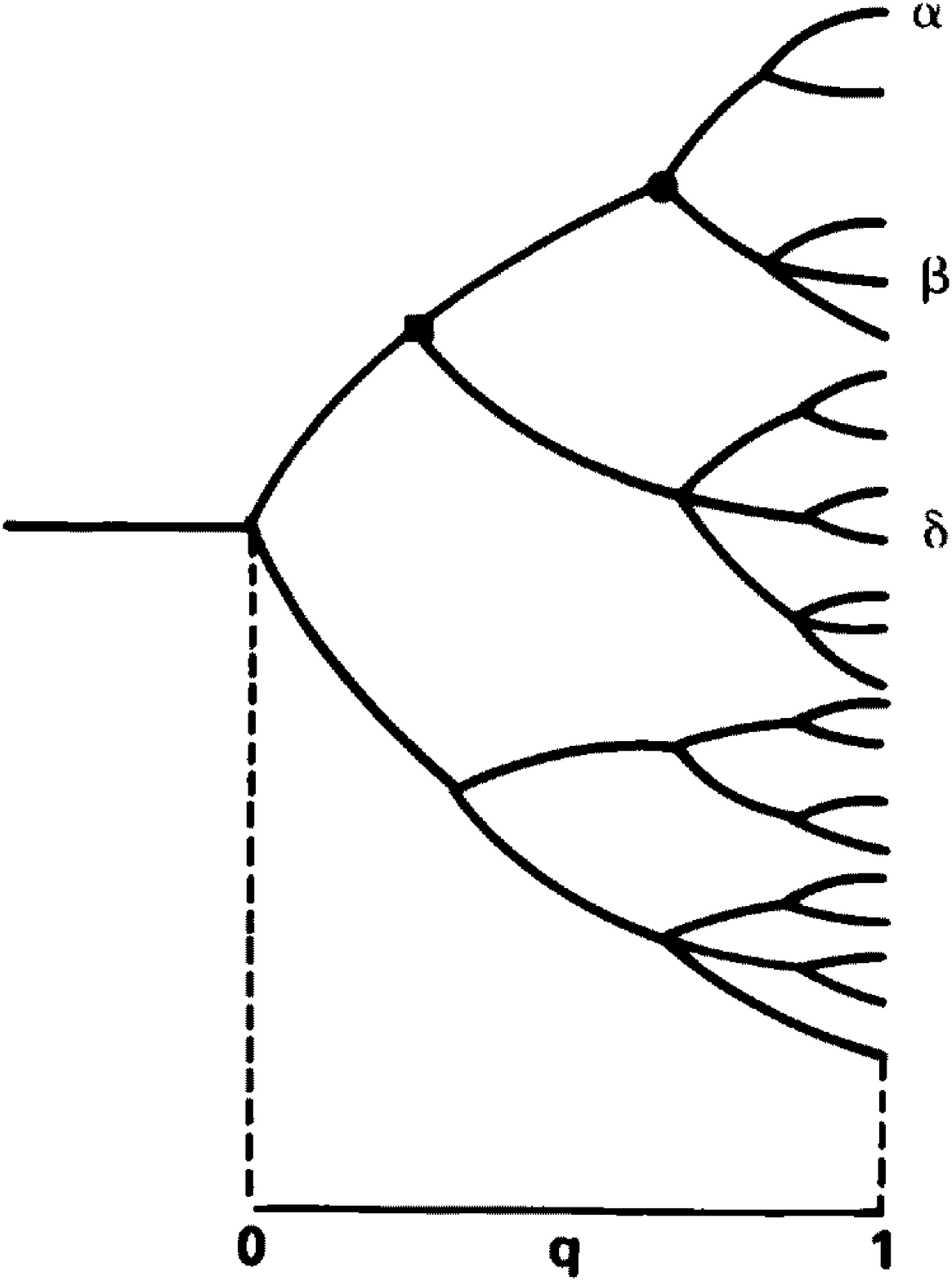}{3in}}
\caption{An evolutionary tree, illustrating the occurrence of ultrametricity.  If the overlap between two final states is measured by the degree of evolution of their nearest common ancestor, as shown, then for any group of three final states, the two smallest overlaps are equal.}
\end{figure}

\subsection{Non-self-averaging}  As discussed earlier, intuition from ergodic-type ideas in conventional statistical mechanics leads one to expect that physical observables will possess the self-averaging feature; i.e. for an observable $A$ of positive expectation value
\begin{equation}
\lim_{N\to \infty} \left\{\frac{\{[A^2] - [A]^2\}^{1/2}}{[A]}\right\} = 0.
\end{equation}
The Parisi Ansatz indeed maintains self-averaging for the normal observables such as energy, free energy and their derivatives.  On the other hand, however, the distribution of overlaps is not self-averaging and therefore neither are its moments.  For example, consider $A=q=N^{-1}\sum_i\langle\sigma_i\rangle^2$.  Then
\begin{equation}
(\Delta q)^2 = [q^2] - [q]^2 = \int^1_0 q^2(x)dx - \left(\int^1_0 q(x) dx \right)^2,
\end{equation}
which is zero if $q(x)$ is flat (replica-symmetric) but not if $q(x)$ has structure; i.e. replica-symmetry breaking leads to non-self-averaging of the overlaps, again a manifestation of the many non-equivalent states.  In fact, within the Parisi Ansatz the overlap probability measure which is self-averaging is the distribution function of the distribution function $P(P(q))$ \cite{MPSTV}.

The full Parisi analysis of the SK model leads to a further deduction that the free-energy landscape evolves chaotically as the parameters are changed; i.e. a small change in the actual set of $\{J_{ij}\}$ or of the temperature or applied field causes a non-trivial change in the hill-valley structure \footnote{This is not the case for all spin glass models.}.\vskip0.10in

\subsection{The Sherrington-Kirkpatrick model}

Although mean-field theory can only be considered approximate for systems with short-range interactions, the above analysis is believed to be exact for a special model in which every spin interacts with every other spin, but via quenched random interactions with an appropriate scaling with $N$.  This is the Sherrington-Kirkpatrick (SK) model \cite{SK} characterized in its Ising version by a Hamiltonian.
\begin{equation}
H = - \sum_{(ij)} J_{ij}\sigma_i\sigma_j,
\end{equation}
where the sum is over all pairs $(ij)$ and the $J_{ij}$ are quenched parameters chosen randomly and independently from a distribution with moments
\begin{equation}
[J_{ij}] = \tilde{J}_0 = J_0/N, ~ ~ ~ ~ 
[J^2_{ij}] = \tilde{J}^2 = J^2/N.
\end{equation}
(\ref{eq:fenergy2}) is exact for this problem with $\tilde{J}_0z \to J_0,\tilde{J}z^{1/2} \to J$.  This is readily seen by the following procedure: (1) replica theory is used to express $[F]$ as in (\ref{eq:fenergy}) but with $(ij)$ now running over all pairs of sites, (2) the $\sigma -$ summations in the exponent are written as complete squares
\begin{equation}
\sum_{(ij)}\sigma^\alpha_i\sigma^\alpha_j = {1\over 2} ((\sum_i\sigma^\alpha_i)^2 -1 ),
\end{equation}
\begin{equation}
\sum_{(ij)}\sigma^\alpha_i\sigma^\beta_i\sigma^\alpha_j\sigma^\beta_j = {1\over 2}((\sum_i\sigma^\alpha_i\sigma^\beta_i)^2 -1),
\end{equation}
and eq. (3.25) used to express $[F]$ in terms of integrals of effective single-site paramagnets over auxiliary fields as
\begin{equation}
[F]=-kT\lim_{N\to\infty}\lim_{n\to 0}\frac{1}{n}\left[\int\Pi_\alpha d\tilde{m}^\alpha\Pi_{(\alpha\beta)}d\tilde{q}^{(\alpha\beta)}\exp(-N\beta g(\{\tilde{m}^\alpha\},\{\tilde{q}^{(\alpha\beta)}\}))-1\right]\label{eq:extrF}
\end{equation}
where
\begin{eqnarray}
g(\{\tilde{m}^\alpha\},\{\tilde{q}^{(\alpha\beta)}\}) = (\tilde{m}^\alpha)^2/2 &+& (\tilde{q}^{(\alpha\beta)})^2/2 - kT~\ln~Tr\exp[\beta J_0\sum_\alpha\sigma^\alpha\tilde{m}^\alpha \nonumber \\
&+&(\beta J)^2(n+2\sum_{(\alpha\beta)}\sigma^\alpha\sigma^\beta\tilde{q}^{(\alpha\beta)})].\end{eqnarray}
Because $g$ is independent of $N$, the integral of eq.(\ref{eq:extrF}) is overwhelmingly dominated by the values of $\{\tilde{m}^\alpha\},\{\tilde{q}^{(\alpha\beta)}\}$ for which $g$ is minimized.  This extremum has
\begin{equation}
\tilde{m}^\alpha = m^\alpha, ~ ~ ~ ~ ~ \tilde{q}^{(\alpha\beta)} = q^{(\alpha\beta)}
\end{equation}
where $m^\alpha,q^{(\alpha\beta)}$ are as given in (\ref{eq:mf3}).  The free energy $[F]$ is as given in (\ref{eq:fenergy2}).  Thus the further analysis of (\ref{eq:mf3}) discussed earlier is believed to apply to the SK model; mean field theory is exact in the limit $N\to\infty$.  Computer simulations of $P(q)$ and of ultrametricity confirm the results obtained analytically for this model \cite{Y,BY}.

\subsection{Other infinite-ranged models}

Other infinite-ranged models are soluble by similar techniques, together with some further transformations, and are taken to define the corresponding mean field solutions.  The general procedure is as follows
\begin{itemize}
\item[(i)] The partition function is given by 
\begin{equation}
Z=Tr_{\{{\bm{S}}\}} \exp(-\beta H(\{{\bm{S}}\})
\end{equation}
where the ${\bm{S}}$ are the microscopic variables.

\item[(ii)] Replicas are used to express $\ln Z$ in a form convenient to average over the quenched disorder
\begin{equation}
\ln Z = \lim_{n\to0}\frac{1}{n}(Z^n-1);
\end{equation}
\begin{equation}
Z^n = Tr_{\{{\bm{S}^\alpha\}}_{\alpha=1...n}} \exp(-\beta\sum_\alpha H(\{{\bm{S}}^\alpha\}))
\end{equation}

\item[(iii)] An average is performed over the quenched disorder to yield
\begin{equation}
[Z^n] = Tr_{\{{\bm{S}}^\alpha\}}\exp(-\beta H_{eff}(\{{\bm{S}}\}))
\end{equation}

\item[(iv)] For infinite ranged problems $H_{eff}$ will always involve the microscopic variables only as functions of sums over all sites; i.e. in combinations $N^{-1}\sum_i{\bm{S}}_i^\alpha,N^{-1}\sum_i{\bm{S}}_i^\alpha{\bm{S}}_i^\beta$.  This is the source of solvability.  Relevance and physical meaningfulness also require appropriate $N-$scaling of the quenched parameters in $H$, such that the resultant effective energy is extensive.

\item[(v)] Macroscopic variables are introduced via the insertion of appropriate representations of unity, for example 
\begin{eqnarray}
&1&=\int d{\bm{m}}^\alpha\delta({\bm{m}}^\alpha - N^{-1}\sum_i{\bm{S}}_i^\alpha) \nonumber\\
&1&=\int d{\bm{q}}^{\alpha\beta}\delta({\bm{q}}^{\alpha\beta} - N^{-1} \sum_i{\bm{S}}_i^\alpha{\bm{S}}_i^\beta)
\end{eqnarray}

These are then used to re-write $H_{eff}$ in terms of macroscopic variables alone.

\item[(vi)] The delta functions are re-written in exponential form; e.g.
\begin{equation}
\delta(q^{\alpha\beta}-N^{-1}\sum_iS_i^\alpha S_i^\beta)=\int d {\bm{\hat{q}}}^{\alpha\beta}\exp(i{\bm{\hat{q}}}^{\alpha\beta}({\bm{q}}^{\alpha\beta}-N^{-1}\sum_i{\bm{S}}^\alpha_i{\bm{S}}^\beta_i))
\end{equation}

The trace over the microvariables now separates so as to yield formally
\begin{equation}
[Z^n]=\int\Pi_\alpha d{\bm{x}}^\alpha\Pi_{\alpha\beta}d{\bm{y}}^{\alpha\beta}\exp(-N\Phi(\{{\bm{x}}^\alpha,{\bm{y}}^{\alpha\beta}\}))
\end{equation}
where ${\bm{x}},{\bm{y}}$ denote all the introduced macrovariables and $\Phi$ involves only a single-site microvariable trace.  Since the number of macrovariables is finite the integral is dominated in the limit of large $N$ by the extremum of $\Phi$. This then yields a set of coupled equations for the extremal ${\bm{x}},{\bm{y}}$, which are themselves identifiable as thermal averages; for example the extremal ${\bm{m}}^\alpha$ is identified with $N^{-1}\sum_i\langle{\bm{S}}_i^\alpha\rangle$.

\item[(vii)] There still remains the problem of the explicit evaluation of $\Phi$ and of taking the limit $n\to0$.  For this the Parisi ansatz is used (or a simpler ansatz if stable).
\end{itemize}

In the above I have written the macrovariables ${\bm{S}}$ in a notation reminiscent of spins and indeed there are many examples in which they are spins or analogues thereof (for example, quadrupoles or Potts spins \footnote{For a review of the situation for vector and Potts spin glasses, but excluding consideration of a recently recognised subtlety for Potts dimension greater than four, the reader is referred to an earlier review by the author \cite{S2}} or states of neurons \footnote{A discussion of applications to neural networks is given in section 5; see also ref \cite{S3}.}), but there are also examples in which the microvariables are more reminiscent of exchange interactions, with the disorder in the consequential spin ordering.  An example of the latter situation occurs in the optimization of neural networks for associative memory recall \cite{S3}; we return to this later.

\subsection{One step replica symmetry breaking}

As noted earlier, the SK model requires the full hierarchy of replica symmetry breakings, implying a complex free energy landscape with hills and valleys on all scales.  Recently, however, interest has turned to some systems which require only a single step of replica symmetry breaking (1RSB) but which turn out to have other subtleties and probable relevance to conventional glasses.  The full picture of these systems is still being developed but one of the important ingredients in the emerging understanding concerns the character of the 1RSB.

In 1RSB the Parisi order function $q(x)$ has the form 
\begin{eqnarray}
q(x) &=& q_0; 0 \leq x \leq m \nonumber \\
& & q_1;m < x \leq 1 ~ ~ ~ ; q_1 \geq q_0.
\end{eqnarray}
whereas in the RS phase $q(x) = q; ~{\rm all}~ x$.

There are two possible scenarios for the onset of 1RSB; in `continuous 1RSB' $(q_1-q_0)$ grows continuouly as the threshold is passed, while in `discontinuous 1RSB' the onset of a non-zero $(q_1-q_0)$ is discontinuous.  In both cases the onset of the anomaly $\Delta=(q_1-q_0)(1-m)$ is continuous, so discontinuous 1RSB commences with a continuous growth of $(1-m)$.  

In terms of the overlap distribution $[P(q)]$ the picture is as follows.  Above the critical temperature (or other control parameter), in the RS region, $[P(q)]$ has a single delta function indicating that the thermodynamics is dominated by a single state.  For continuous 1RSB a second delta-function peak grows continuously away from the first as the transition temperature is crossed, the higher-$q$ peak having a finite weight $(1-m)$, the lower a weight $m$. For the onset of discontinuous 1RSB again there appears a second delta function peak but at a discontinuously larger value of $q$.  This higher-$q$ peak starts with a small weight $(1-m)$ which grows continuously as the temperature is lowered, with relatively much less change in the value of $q$.

These observations suggest rather different images for the evolutions of free energy landscapes in  systems with continuous and discontinuous RSB, both different from that of an RS system.  The transition from paramagnet to RS order is envisaged as one in which the free energy evolves from a state with a single paramagnetic valley to one with a finite number of ordered states which are equivalent via global symmetry transformations (eg spin inversion).  For a continuous transition these ordered states grow smoothly from the paramagnetic state.  For continuous RSB the transition is still continuous but to a thermodynamically relevant number \footnote{i.e. a number exponential in $N$.} of non-equivalent ordered states.  For 1RSB these states are equally mutually orthogonal and their self-overlap grows continuously from their mutual overlap as the temperature is lowered; the mutual overlap is zero in the absence of a field.  For full RSB (as in the SK model) these ordered states have a continuous range of mutual overlaps.  For discontinuous 1RSB the ordered states appear with thermodynamic relevance already with large self-overlaps, and again with equal mutual overlaps, but their number grows continuously as the temperature is lowered.  One is therefore led to the conception that these states already exist above the transition temperature, with corresponding valleys in the energy landscape, but that their free energy relevance only becomes dominant when the transition temperature is reached and increases as the temperature is lowered.  This already gives a hint also that their dynamical behaviour above the glass temperature might be different from systems with continuous RSB.

One example of a system which exhibits 1RSB is the p-spin spherical spin glass \cite{CS}, whose Hamiltonian is given in (\ref{eq:pspin}).  For fields less than a critical value $h_c$ the onset of 1RSB is discontinuous, for $h>h_c$ it is continuous.  Other examples are found in Potts and quadrapolar glasses for high enough field dimension \cite{S2}.  Similar but self-induced behaviour is believed to control the glass transition in fragile glasses \cite{BCKM}.

\subsection{Further subtleties}

Before leaving this replica analysis some further subtleties are worthy of note since they illustrate the need for caution in passing from conventional procedures to unconventional phase transition problems.

The extremum of $[F]$ giving the phase correctly is that {\it maximizing} $[F]$ with respect to $q$, in stark contrast to conventional minimization.  This is a consequence of the fact that the number of $(\alpha\beta)$-combinations, $n(n-1)/2$, becomes negative as $n$ tends to zero.

A second unconventional-looking feature is that if one analytically continues the high temperature free energy $[F]$, with $m=q=0,$ to low temperature, the result is lower than the free energy of the spin glass solution.  In fact, however, the analytic continuation of a state beyond the limit of its validity is strictly meaningless.

Within RSB theory there are several parameters against which to extremize and the most obvious choice is not always correct; for example in 1RSB one is tempted to extremize with respect to $q_0,q_1$ and the break point $m$.  Although this works for continuous 1RSB, for discontinuous 1RSB it gives results different from those obtained from dynamics.

In fact the relevant criteria are that systems must be stable against all fluctuations and that configurational entropies must also be taken into account.

\section{Dynamics}

Thus far our discussion has concerned thermodynamics, analyzed from the perspective of the partition function as generator, albeit probed to yield insight more deeply into the metastable/pure state structure.  Such analysis can be taken (and is still being taken) further to probe the state structure more deeply, for example by imposing overlaps between real replicas or by employing non-Boltzmann weights.

However, often one would like to study dynamics.  This is clearly necessary to learn about properties away from equilibrium, but it is not clear that spin glass systems do equilibrate.  In the Introduction some experimental and simulational observations of preparation dependence and slow response were mentioned.  In fact there are many more, of which one of the most intriguing is the phenomenon of aging, where the response of a system depends on how long it has been waiting since preparation.  Hence, it is interesting to discuss dynamics, allowing for the system to be away from equilibrium. \footnote{To aid this distinction the statistical mechanics of equilibrium is now often referred to the spin glass literature as `statics'.}  Recent studies have again exposed more interesting observations, implications, applications and conceptual challenges than could have been anticipated.

Non-equilibrium dynamics is also a subject of considerable current interest in other areas of statistical physics, where systems without detailed balance in their macroscopic dynamics, and therefore no Boltzmann-like equilibrium even if a steady state is reached, show dynamically induced phase transitions.  Here, however, I shall restrict discussion to spin-glass like systems.

\subsection{Microscopic dynamics}

The systems we consider are describable by Markovian microscopic dynamics; i.e. local in time and stochastic.  Examples are
\begin{itemize}
\item[(i)] Ising spins satisfying either random sequential or parallel Glauber dynamics in which spins are updated with probabilities
\begin{equation}
{\rm Prob}[\sigma_i\to\sigma^\prime_i] = {1\over2}[1 + \tanh(\beta h_i\sigma_i^\prime)]
\end{equation}
where $h_i$ is the effective field experienced by spin $i~ (h_i=\sum_jJ_{ij}\sigma_j$ for the Hamiltonian of (\ref{eq:sgH})).

\item[(ii)] Metropolis dynamics in which spins are chosen randomly and flipped if either the flip would lower the energy or with probability $\exp(-\Delta E/T)$ if the energy change $\Delta E$ is positive.\footnote{Now we choose units with $k=1$.}

\item[(iii)] Soft spins obeying Langevin dynamics 
\begin{equation}
\frac{\partial\phi_i}{\partial t} = \frac{-\partial H}{\partial\phi_i} + \eta_i(t)
\end{equation}
where $\eta_i(t)$ is white noise of variance $T$.
\end{itemize}  

We might note in passing that for a system to yield a thermodynamic equilibrium distribution of the Boltzmann type it is necessary for the dynamics to obey appropriate stochastic balance conditions \cite{van Kampen,Parisi}.  These are satisfied by (i) - (iii) with random sequential dynamics, symmetric interactions $(J_{ij} = -J_{ji})$ and no self-interaction $(J_{ii}=0)$.  For parallel dynamics a different but also straightforward equilibrium distribution results \cite{Peretto}.

\subsection{Macroscopic dynamics}

There are several ways to study the macroscopic dynamics.  One is to try to use the microscopic equations to study directly the evolution of specified macroscopic parameters.  In general the equation of motion of one such parameter leads to a dependence on a higher one, whose equation of motion leads to still more in a never-terminating sequence.  Well chosen order parameters and Ans\"atze can yield closure but usually only approximately.  In a later section we shall introduce one such procedure which appears to work quite well in comparisons with simulation.

A second is to employ a generating functional in analogy with the partition function for thermodynamics.  In this section I shall outline this procedure applied to a system with infinite-ranged interactions.

The underlying principle is to introduce and study a generating functional
\begin{equation}
Z_{\{{\bm{\lambda}}(t)\}}[\{{\bm{S}}(t)\}]= Tr_{\{{\bm{S}}(t)\}}\delta({\rm Micros.~eqn~of~motion)}\exp(\int dt{\bm{\lambda}}(t){\bm{S}}(t))
\end{equation}
where $\delta$(Micros. eqn. of motion) is a shorthand for the restriction of the microscopic variables to ones obeying the microscopic equations of motion.  If necessary a Jacobian is also included to ensure $Z_{\{{\bm{\lambda}}(t)=1\}}[\{{\bm{S}}(t)\}]=1$.  Then any desired correlation or response functions may be obtained as derivatives, for example
\begin{eqnarray}
C_{ij}(t,t^\prime)&=& \langle {\bm{S}}_i(t){\bm{S}}_j(t^\prime)\rangle\\
&=&\lim_{{\bm{\lambda}}\to0}\frac{\partial}{\partial{\bm{\lambda}}_i(t)}\frac{\partial}{\partial{\bm{\lambda}}_j(t^{\prime)}}\ln Z_{\{{\bm{\lambda}}\}}\\
&=&\lim_{{\bm{\lambda}}\to0}\frac{\partial}{\partial{\bm{\lambda}}_i(t)}\frac{\partial}{\partial{\bm{\lambda}}_j(t)^{\prime)}}Z_{\{{\bm{\lambda}}\}}
\end{eqnarray}
The last identity follows from the restriction $Z_{\{\lambda=1\}}=1$.  This ensures also that any disorder averaging can be performed directly on $Z$ (without need for replicas).  Sums of microscopic variables are then eliminated in favour of macroscopic variables by a dynamic analogue of the procedure of section (3.4) to yield a description in terms of an integral over macroscopic variables alone.  For infinite-ranged models the integral is extremally dominated and extremization yields coupled equations for macroscopic order parameters, correlation and response functions.

Let us consider the process slightly more explicitly for a system of soft spins obeying Langevin dynamics, for example the p-spin spherical spin glass model \cite{CSH} in which \begin{equation}
H=-\sum_{i_{1}<i_{2}...<i_{p}}J_{i_{1}i_{2}...i_{p}}\phi_{i_{1}}\phi_{i_{2}}\cdots\phi_{i_{p}} - \sum_ih_i\phi_i,\label{eq:pspin}
\end{equation}
the $J$ are chosen independently randomly from a Gaussian distribution of variance $J^2p!/2N^{p-1}$ and the $\phi$ satisfy the spherical constraint $\sum\phi^2_i=N$.  The constraint can be implemented conveniently by adding a chemical potential term $\mu(t)\sum\phi^2_i$ to $H$ and determining $\mu(t)$ self-consistently.  For simplicity we shall discuss explicitly only $h=0$.

Before averaging over the stochastic noise
\begin{equation}
Z=\int\delta\phi\Pi_{it}\delta(\frac{\partial}{\partial t}\phi_i(t)  + \frac{\partial H}{\partial\phi_i(t)} - \eta_i(t)).
\end{equation}
Exponentiating the $\delta$ functions introduces a new set of microvariables 
\begin{equation}
Z=\int\delta\phi\delta\hat{\phi}\exp\left\{\sum_{it}\left[\hat{\phi}_i(t)(\frac{\partial\phi_i(t)}{\partial t} + \frac{\partial H}{\partial\phi_i(t)} - \eta_i(t))\right]\right\}
\end{equation}
Averaging over the stochastic noise, which is assumed to be Gaussian distributed and local in $i,t$ with variance $T$, yields 
\begin{equation}
\langle Z\rangle = \int\delta\phi\delta\hat{\phi}\exp\left\{\sum_{it}\left[\hat{\phi}_i(t)(\frac{\partial\phi_i(t)}{\partial t} + \frac{\partial H}{\partial\phi_i(t)})+T\hat{\phi}_i(t)\hat{\phi}_i(t)    \right]     \right\}
\end{equation}

The quenched disorder enters through terms of the form
\begin{equation}
\exp\left\{\sum_t\sum_{i_{1}...i_{p}}J_{i_{1}...i_{p}}\tilde{\phi}_{i_{1}}(t)\cdots\tilde{\phi}_{i_{p}}(t) \right\}
\end{equation}
where one of the $\tilde{\phi}$ is $\hat{\phi}$ and the others are $\phi$.  Averaging over the disorder yields
\begin{eqnarray}
[\langle Z\rangle] &=& \int\delta\phi\delta\hat{\phi}\exp\{\sum_{it}[\hat{\phi}_i(t) \{\frac{\partial\phi_i(t)}{\partial t} + \mu\phi_i(t)\}+T\hat{\phi}_i(t)\hat{\phi}_i(t)] \nonumber \\
&+&\beta^2J^2\sum_{t,t^{\prime}}\sum_{{\rm comb}} (\sum_{i_{1}}\tilde{\phi}_{i_{1}}(t)\tilde{\phi}_{i_{1}}(t^\prime))\cdots (\sum_{i_{p}}\tilde{\phi}_{i_{p}}(t)\tilde{\phi}_{i_{p}}(t^\prime))]\}
\end{eqnarray}
where $\sum_{{\rm comb}}$ means a sum over all combinations in which one $\tilde{\phi}(t)$ and one $\tilde{\phi}(t^\prime)$ are $\hat{\phi}(t),\hat{\phi}(t^\prime)$ respectively and all the other $\tilde{\phi}$ are $\phi$.

It will be noted that all the microscopic variables only enter in the macroscopic form $\sum_i\tilde{\phi}_i(t)\tilde{\phi}_i(t^\prime)$ and hence they can be eliminated by use of
\begin{eqnarray}
&1&=\int\delta C(t,t^\prime)\delta(C(t,t^\prime) - N^{-1}\sum_i\phi_i(t)\phi_i(t^\prime)) \nonumber \\
&1&=\int\delta G(t,t^\prime)\delta(G(t,t^\prime) - N^{-1}\sum_i\phi_i(t)\hat{\phi}_i(t^\prime)) \nonumber \\
&1&=\int\delta D(t,t^\prime)\delta(D(t,t^\prime) - N^{-1}\sum_i\hat{\phi}_i(t)\hat{\phi}_i(t^\prime)).
\end{eqnarray}

Further exponentiation of these $\delta$-functions via auxiliary macrofunctions $\hat{C}(t,t^\prime),\hat{G}(t,t^\prime), \hat{D}(t,t^\prime)$ yields separable local Gaussian integrals in $\phi, \hat{\phi}$, which are readily performed to yield
\begin{equation}
[\langle Z\rangle]=\int\delta C\delta\hat{C}\delta G\delta\hat{G}\delta  D\delta\hat{D}\exp(-N\Phi(C,\hat{C},G,\hat{G},D,\hat{D}));
\end{equation}

Since the present purpose is only to illustrate procedures, and to save space, the explicit form of $\Phi$ is not given here.  A similar calculation can be applied to other soft-spin models, resulting in different $\Phi$, but the following remarks continue to hold.  Provided that $N\gg t,t^\prime$ the integral is dominated by the extremum which yields a set of coupled equations for the macroscopic order functions (of which only $C(t,t^\prime)$ and $G(t,t^\prime);t^\prime < t$ are non-zero) \footnote{The vanishing of $D$ and $G(t,t^\prime);t^\prime>t$ corresponds to causality.}.

These extremal order functions can be identified with the correlation and response functions

\begin{eqnarray}
&C&(t,t^\prime) = N^{-1}\sum_i\langle\phi_i(t)\phi_i(t^\prime)\rangle \nonumber \\
&G&(t,t^\prime) = N^{-1}\sum_i\langle\phi_i(t)\hat{\phi}_i(t^\prime)\rangle\equiv N^{-1}\sum_i\partial\langle\phi_i(t)\rangle/\partial h_i(t^\prime)|h_{i=0}
\end{eqnarray}

$C(t,t^\prime)$ and $G(t,t^\prime)$ satisfy coupled integro-differential equations, non-local in time.  For the p-spin spherical model these are

\begin{eqnarray}
\frac{\partial}{\partial t} G(t,t^\prime)=&&\delta(t-t^\prime) - \mu(t)G(t,t^\prime) \nonumber \\
&+&{1\over 2}p(p-1)\beta^2J^2\int^t_{t^{\prime}}ds G(t,s)C^{p-2}(t,s)G(s,t^\prime) \nonumber \\
\frac{\partial}{\partial t}C(t,t^\prime)=&-&\mu(t)C(t,t^\prime) + 2G(t^\prime,t) \nonumber \\
&+& {1\over 2}p(p-1)\beta^2J^2\int^t_0ds G(t,s)C^{p-2}(t,s)C(s,t^\prime) \nonumber \\
&+& {1\over 2}p\beta^2J^2\int^{t^{\prime}}_0dsC^{p-1}(t,s)G(t^\prime,s)
\end{eqnarray}

Above a dynamical transition temperature $C$ and $G$ are time-translational invariant (TTI), dependent only on the relative time $\tau$.  They obey the usual fluctuation-dissipation theorem $G_{FDT}(\tau) = - \theta(\tau)\partial_\tau C_{FDT}(\tau)$.  For the p-spin spherical model, and other models with discontinuous 1RSB in their statics, they also lead to mode-coupling equations similar to those proposed for glasses \cite{G}, with consequential fast $\beta$-relaxation to a plateau followed by slow $\alpha$-relaxation;
\begin{eqnarray}
&\beta&-{\rm relax}: C(\tau)\sim q_{EA} + C_\beta \tau^{-\beta}; C \geq q_{EA} \nonumber \\
&\alpha&-{\rm relax}: C(\tau)\sim q_{EA} - C_\alpha\tau^{-\alpha}; C \leq q_{EA}.
\end{eqnarray}
As the temperature is lowered, the plateau region grows until $C(\tau)$ no longer decays to zero but remains a long time at a finite value $q_{EA}$, corresponding to the $q_1$, which appears discontinuously at the static transition.   This defines a dynamical transition temperature\footnote{This temperature is higher than that obtained by naive replica theory; a more sophisticated replica analysis based on marginal stability rather than extremization against $m$, or taking correct account of configurational entropy, gives the correct transition temperature.}.  For models whose statics has continuous RSB $C(\tau)$ decays to zero, but with critical slowing down, at the replica theory transition temperature and the plateau $q_{EA}$ grows continuously as the temperature is lowered further.

Beneath the transition temperature a new behaviour is found in that both TTI and FDT break down at long relative times and aging is observed.  This shows up for example in $C(\tau+t_w,t_w)$, where the times are measured from the time of preparation of the system at the temperature of study, after a quench from a high temperature state.  When the relative time $\tau$ is much less than the waiting time $t_w$ TTI and FDT hold and $C$ settles to a plateau value $q_{EA}$, corresponding to the self-overlap peak in the static $P(q)$.  However, when $\tau$ becomes of the order of $t_w ~ ~ C(\tau+t_w,t_w)$ starts to decay again, reducing asymptotically towards zero.  The phenomenon of dependence on $t_w$ is known as `aging', the slow decay to zero as `weak ergodicity breaking'. Thus \\ $C(\tau+t_w,t_w)$ is often expressed in terms of two parts

\begin{equation}
C(\tau+t_w,t_w) = C_{ST}(\tau) + C_{AG}(\tau+t_w,t_w)
\end{equation}
where $C_{ST}(\tau)$ is TTI and satisfies FDT, whereas $C_{AG}$ is the aging contribution.  $C_{ST}(\tau)$ starts from its maximum value and decays rapidly towards a plateau at $q_{EA}$.  $C_{AG}$ is zero for small $\tau$ and goes asymptotically towards $(-q_{EA})$ on a  time-scale given by $t_w$; for the p-spin spherical model the decay is as~\footnote{A more general form is as $\sum_\lambda C_\lambda[h_\lambda(t_w)/h_\lambda(\tau+t_w)]$} $(t_w/(\tau+t_w)^\gamma$.

In the low temperature region there is a modified FDT

\begin{equation}
G(t,t^\prime) = \frac{X(t,t^\prime)}{T} \frac{\partial C(t,t^\prime)}{\partial t^\prime}.
\end{equation}
At least for the p-spin spherical model $X$ appears to depend on the times only via $C$

\begin{equation}
G(t,t^\prime)=\frac{X[C(t,t^\prime)]}{T} \frac{\partial C(t,t^\prime)}{\partial t^\prime}.
\end{equation}
Thus the system behaves as though it has an effective temperature $T/X$.

Several of these properties are reminiscent of real glasses \cite{G,Struick} and so similar procedures are currently being investigated for such systems, with rapid freeze-out of a glassy configuration providing the effective disorder controlling the remaining degrees of freedom.

For a more extensive review of dynamics away from equilibrium the reader is referred to \cite{BCKM}

\section{Beyond conventional spin glasses}

As indicated in the Introduction, concepts and techniques developed for spin glasses are now finding application and extension much more widely in the science of complex cooperative behaviour in disordered (or quasi-disordered) and frustrated many-body systems.  In this section two such areas of application are introduced, namely neural networks and hard optimization.

\subsection{Neural networks: the spin glass approach}

The relevance of conventional spin glasses to neural networks lies not in any physical similarity, but rather in conceptual analogy and in the transfer of mathematical techniques developed for the analysis of spin glasses to the quantitative study of several aspects of neural networks.  This section is concerned with the basis and application of this transfer.

Just as for spin glasses, neural networks also involve the cooperation of many relatively simple units, the neurons, under the influence of conflicting interactions, and they possess many different global asymptotic behaviours in their dynamics.  In this case the conflicts arise from a mixture of excitatory and inhibitory synapses, respectively increasing and decreasing the tendency of a post-synaptic neuron to fire if the pre-synaptic neuron fires.

The recognition of a conceptual relationship between spin glasses and recurrent neural networks, together with a mathematical mapping between idealizations of each \cite{H}, provided the first hint of what has turned out to be a fruitful transplantation.

In fact, there are now several respects in which spin glass analysis has been of value in considering neural networks for storing and interpreting static data.  One concerns the macroscopic asymptotic behaviour of a neural network of given architecture and synaptic efficacies.  A second concerns the choice of efficacies in order to optimize various performance measures.  A third concerns the dynamics of associative recall, learning and generalization.

We shall discuss networks suggested as idealizations of neurobiological stuctures and also those devised for applied decision making.  We shall not, however, dwell on the extent to which these idealizations are faithful, or otherwise, to nature.

\subsection{Types of Neural Network}

There are two principal types of neural network architecture which have been the subject of active study.  

The first is that of layered feedforward networks in which many input neurons drive various numbers of hidden units eventually to one or few output neurons, with signals progressing only forward from layer to layer, never backwards or sideways within a layer.  This is the preferred architecture of many artificial neural networks for application as expert systems, with the interest lying in training and operating the networks for the deduction of appropriate few-state conclusions from the simultaneous input of many, possibly corrupted, pieces of input data.

The second type is of recurrent networks where there is no simple feedforward-only or even layered operation, but rather the neurons drive one another collectively and repetitively without particular directionality.  In these networks the interest is in the global behaviour of all the neurons and the associative retrieval of memorized states from initializations in noisy representations thereof.  These networks are often referred to as {\it attractor} neural networks \footnote{They are often abbreviated as ANN, but we shall avoide this notation since it is also common for {\it artificial} neural networks}.  They are idealizations of parts of the brain, such as cerebral cortex.  Historically, they were the stimulation for devising artificial networks.  

Both of the above can be considered as made up from simple `units' in which a single neuron received input from several other neurons which collectively determine its output.  That output may then, depending upon the architecture considered, provide part of the inputs to other neurons in other units.

Many specific forms of idealized neuron are possible, but here we shall concentrate on those in which the neuron state (activity) can be characterized by a single real scalar.  Similarly, many types of rule can be envisaged, relating the output state of a neuron to those of the neurons which input directly to it.  We shall concentrate, however, on those in which the efferent (post-synaptic) behaviour is determined from the states of the afferent (pre-synaptic) neurons via an `effective field'
\begin{equation}
h_i = \sum_{j\not=i} J_{ij}\phi_j-W_i,
\end{equation} 
where $\phi_j$ measures the firing state of neuron $j$, $J_{ij}$ is the synaptic weight from $j$ to $i$, and $W_i$ is a threshold.  For example, a deterministic perceptron obeys the output-input relation
\begin{equation}
\phi^\prime_i = f(h_i),
\end{equation}
where $\phi_i$ is the output state of the neuron and $f(h)$ is a non-linear sigmoidal function.

It is  often convenient  technically to  specialize further to binary-state\\ (McCulloch-Pitts) neurons, taken to have $\phi_i \to \sigma_i = \pm 1$ denoting firing/non-firing.  One cannot then have a deterministic sigmoidal response but instead one can interpret the sigmoidal function, suitably normalized, as a measure of the {\it probability} of firing.  In the spirit of picking out the essentials it is convenient to employ a sigmoid function characterized by a single rounding parameter.  One such choice is the Glauber rule.
\begin{equation}
\sigma_i \to \sigma^\prime_i ~{\rm with~probability}~ {1\over 2}[1 + \tanh(\beta h_i\sigma^\prime_i)],
\end{equation}
where now $T=\beta^{-1}$ is a measure of the degree of stochasticity or rounding of the initial sigmoidal response function, with $T=0~(\beta=\infty)$ corresponding to determinism.  In a network of such units, updates can be effectuated either synchronously (in parallel) or randomly asynchronously; we shall concentrate on the latter which is more analagous to a spin glass.

Thus we have a system reminiscent of the Ising spin glass model discussed earlier.  The similarities are that neuron states are analagous to the microscopic spin states and the synaptic efficacies are analagous to the exchange interactions .  We expect that at high $T$ both will be effectively paramagnetic, with unconfined dynamics.  Our experience with spin glasses tells us that frustrated interactions lead to a low temperature free energy surface with many valleys, which immediately suggests that a similar frustration in synaptic efficacies will be needed to produce many global attractors for the neural dynamics, to act as memory storages.  Hence the need for both excitatory and inhibitory synapses; $J_{ij}$ both positive and negative.  We shall see that the memories are related to global microstates.  The desire that the dynamics is driven to those microstates implies that for neural networks the $J_{ij}$ must be trained or tuned to put the `valleys' in the correct places; this is the first difference from spin glasses.  A second is that these valleys should have reasonably large basins of attraction, subject to the constraints imposed by having many attractors.  A third is that our interest in neural networks is strictly in the attractor space for the dynamics rather than the thermodynamics of a Boltzmann-weighted Hamiltonian.

Thus we arrive at a mental image of a recurrent neural network as a system of many neurons in which patterns are characterized by particular global microstates ${\bm{\sigma}} = {\bm{\xi}}^\mu;{\bm{\sigma}} = \sigma_1, \sigma_2 ...\sigma_N;\mu=1\cdots p$, where the $\mu$ label the patterns, and which is capable of retrieving those patterns associatively from a noisy start ${\bm{\eta}}^\mu;\eta_i^\mu=c_i\xi^\mu_i$ where $c_i=\pm 1$ with a probability $(1-\tilde{p},\tilde{p})$ where $\tilde{p}$ characterizes the initial noise in the input information.

A potential scenario is to choose a system in which the $J_{ij}$ are such that, beneath an appropriate temperature (stochasticity) $T$, there are $p$ disconnected basins, each having a macroscopic overlap \footnote{A precise definition of overlap is given later in (\ref{eq:overlapnn}).  With the normalization used there an overlap is macroscopic if it is of order 1.} with just one of the patterns and such that if the system is started in a microstate which is a noisy version of a pattern it will iterate towards a distribution with a macroscopic overlap with that pattern and perhaps, for $T \to 0$, to the pattern itself.\footnote{For a network to iterate at $T=0$ precisely to a desired pattern requires an optimized design of the $\{J_{ij}\}$.  This is not necessarily the best design for retrieval at $T\not=0$$^{33}$}  To store many non-equivalent patterns clearly requires many non-equivalent basins and therefore competition among the synaptic weights $\{J_{ij}\}$\footnote{Note that this concept applies even if there is no Lyapunov or energy function.  The expression `basin' refers to a restricted microscopic phase space of the $\{\sigma\}$, even in a purely dynamical context.}. 

The mathematical machinery devised to study ordering in random magnets is thus a natural choice to consider for adaptation for the analysis of retrieval in the corresponding neural networks.  An introduction to this adaptation is the subject of the next section.  However, before passing to that analysis a further analogy and stimulus for mathematical transfer will be mentioned.

This second area for transfer concerns the choice of $\{J_{ij}\}$ to achieve a desired network performance.  Provided that performance can be quantified, the problem of choosing the optimal $\{J_{ij}\}$ is equivalent to one of minimizing some effective energy function in the space of all $\{J_{ij}\}$.  The performance requirements, such as which patterns are to be stored and with what quality, impose `costs' on the $J_{ij}$ combinations, much as the exchange interactions do on a specific choice of the spins in (\ref{eq:sgH}), and there are normally conflicts in matching local with global optimization.  Thus, the global optimization problem is conceptually isomorphic with that of finding the ground state of a spin glass and a conceptual and mathematical transfer has proved valuable.

\subsection{Statistical Physics of Retrieval}

In this section we consider the use of spin glass techniques for the analysis of the retrieval properties of simple recurrent neural networks.

Let us consider such a network of $N$ binary-state neurons, characterized by state variables $\sigma_i=\pm1,i=1...N$, interacting via stochastic synaptic operations and storing, or attempting to store $p$ patterns $\{\xi^\mu_i\} = \{\pm1\};\mu = 1,...p$.  Interest is in the global state of the network.  Its closeness to a pattern can be measured in terms of the corresponding (normalized) overlap
\begin{equation}
m^\mu=N^{-1}\sum_i\xi^\mu_i\sigma_i.\label{eq:overlapnn}
\end{equation}

To act as a retrieving memory the phase space of the system must separate so as to include effectively non-communicating sub-spaces each with macroscopic $O(1)$ overlap with a single pattern.

\subsection{The Hopfield model}

A particularly interesting example for analysis was proposed by Hopfield \cite{H}.  It employs symmetric synapses $J_{ij} = J_{ji}$, no self-interaction $(J_{ii}=0)$ and randomly asynchronously updating dynamics, leading to the asymptotic activity distribution (over {\it{all}} microstates)

\begin{equation}
p({\bm{\sigma}}) \sim \exp ((-\beta E({\bm{\sigma}}));\beta =T^{-1}
\end{equation}
where $E({\bm{\sigma}})$ has the form of (\ref{eq:sgH}).  This permits the applications of the machinery of equilibrium statistical mechanics~\footnote{This is the reason for the interest in employing symmetric synapses even though this is not a requirement for neural storage nor universal biological reality.} to study retrieval behaviour and develop conceptual understanding and quantification.  In particular, one studies the resultant thermodynamic phase structure with particular concern for the behaviour of the $m^\mu$.

Hopfield further proposed the simple synaptic form \footnote{Note that this is infinite-ranged; hence (possibly sophisticated) mean field theory will be applicable.}
\begin{equation}
J_{ij}=N^{-1}\sum_\mu\xi^\mu_i\xi^\mu_j(1-\delta_{ij}),\label{eq:Hopfield}
\end{equation}
inspired by the observations of Hebb.  Let us turn to the analysis and implications of this choice, with all the thresholds $\{W_i\}$ taken to be zero and for random uncorrelated patterns ${\bm{\xi}}^\mu$.

For a system storing just a single pattern, the problem transforms immediately, under $\sigma_i\to\sigma_i\xi_i$, to a pure ferromagnetic Ising model with $J_{ij}=N^{-1}$.  The solution is well known and $m$ satisfies the self-consistency equation
\begin{equation}
m=\tanh(\beta m),
\end{equation}
with the physical solution $m=0$ for $T > 1~ ~(\beta<1)$ and a symmetry-breaking phase transition to two separated solutions $\pm|m|$, with $m\not= 0,$ for $T<1$.

For general $p$ one may express $\exp(-\beta E({\bm{\sigma}}))$ for the Hopfield-Hebb model in a separable form.
\begin{eqnarray}
\exp(-\beta E({\bm{\sigma}})) = \exp[+(\beta/2N)\sum^p_{\mu=1}(\sum^N_{i=1}\xi^\mu_i\sigma_i)^2-\beta p/2] \nonumber \\
=\int\Pi^p_\mu d\tilde{m}^\mu(\frac{\beta N}{2\pi})^{\frac{1}{2}}\exp[\sum^p_{\mu=1}(-N\beta(\tilde{m}^\mu)^2/2 - \beta\tilde{m}^\mu\sum_i\sigma_i\xi^\mu_i)-\beta p/2], \label{eq:HH}
\end{eqnarray}
where we have employed (\ref{eq:squares}).

Because the only term in $\sigma_i$ in (\ref{eq:HH}) has the separable form $\Pi_i\exp(-a_i\sigma_i)$ the sum on $\{\sigma_i\}$ in $Z$ is now straightforward and yields
\begin{equation}
Z=\int(\Pi_\mu d\tilde{m}^\mu(\beta N/2\pi)^{\frac{1}{2}})\exp (-N\beta f(\{\tilde{m}^\mu\}))\label{eq:Zsmallp}
\end{equation}
\begin{equation}
{\rm{where}}~ f(\{\tilde{m}^\mu\})=\sum^p_{\mu=1}(\tilde{m}^\mu)^2/2-(N\beta)^{-1}\sum_i\ln[2\cosh\beta\sum^p_{\mu=1}m^\mu \xi^\mu_i]+p/2N.\label{eq:fsmallp}
\end{equation}\vskip0.10in

\noindent
{\it{Intensive numbers of patterns:}}

If $p$ is finite (does not scale with $N)$, $f(\{\tilde{m}^\mu\})$ is intensive in the thermodynamic limit $(N\to\infty)$ and the integral is extremally dominated.  The minima of $f$ give the self-consistent solutions which correspond to the stable dynamics, although only the lowest minima are relevant to a thermodynamic calculation.  The self-consistency equations are 
\begin{equation}
\tilde{m}^\mu=N^{-1}\sum_i\xi^\mu_i\tanh(\beta\sum_\nu\tilde{m}^\nu\xi_i^\nu) = m^\mu
\end{equation}
yielding the results:~\footnote{The results are quoted for the limit $N\to\infty$.  For finite but large $N$ read\\ $``\not=0" \equiv~``O(1)", ~ ~``0"\equiv~``<O(1)"$.}

\noindent 1) $T=0$: (i) All embedded patterns ${\bm{\xi}}^\mu$ are solutions; ie. there are $p$ solutions (all of equal probability)

 $m^\mu=1;{\rm~one~}\mu
 ~ ~ ~ ~ ~ ~ ~ ~ ~ ~ ~ =0:{\rm~rest~}$

\noindent(ii) There are also mixed solutions in which more than one $m^\mu$ is non-zero.
Solutions of type (i) correspond to memory retrieval, while hybrids of type (ii) are normally not wanted.

\noindent
2) $0.46 > T>0$: There are solutions\\
(i) $m^\mu = m\not=0; {\rm~one~} \mu
~ ~  ~ ~ ~ ~ ~  ~ ~ ~ = 0;$ rest

\noindent (ii) mixed states.\\
\noindent In the case of type (i) solutions, we may again speak of retrieval, but now it is imperfect.

\noindent
3) $1>T>0.46$: Only type (i) solutions remain, each equally stable and with extensive barriers.

\noindent
4) $T>1$: Only the paramagnetic solution (all $m^\mu =0$) remains.

Thus we see that retrieval noise can serve a useful purpose in eliminating or reducing spurious hybrid solutions in favour of unique retrieval.

\subsection{Extensive numbers of patterns:} The analysis of the last section shows no dependence of the critical temperatures on $p$.  This is correct for $p$ independent of $N$ and $N \to \infty$.  However, even simple signal-to-noise arguments demonstrate that interference between patterns will destroy retrieval, even at $T=0$, for $p$ large enough and scaling appropriately with $N$.  Geometrical, information-theoretical and statistical-mechanical arguments in fact show that the maximum pattern storage allowing retrieval scales as $p=\alpha N$, where $\alpha$ is an $N$-independent storage capacity.  Thus we need to be able to analyse retrieval for $p$ of order $N$, which requires a different method than that used in (\ref{eq:Zsmallp}),(\ref{eq:fsmallp}).  One is available from the replica theory of spin glasses.

As noted earlier, physical quantities of interest are obtained from $\ln Z$.  This will depend on the specific set of $\{J_{ij}\}$, which will itself depend on the patterns $\{\xi^\mu\}$ to be stored.  Statistically, however, one is interested not in a particular set of  $\{J_{ij}\}$ or $\{\xi^\mu_i\}$ but in relevant averages over generic sets, for example over all sets of $p$ patterns drawn randomly from the $2^N$ possible pattern choices.  Furthermore, the pattern averages of most interest are self-averaging, strongly peaked around their most probable values.  Thus, we may consider $\langle\ln Z\rangle_{\{\xi\}}$ where $\langle ~ \rangle_{\{\xi\}}$ means an average over the specific pattern choices.

The methodology is analgous to that employed in our replica study of spin glasses.  In place of the ferromagnetic order parameter $m$ one now has all the overlap parameters $m^\mu$.  However, since we are principally interested in retrieving symmetry-breaking solutions, we can concentrate on extrema with only one, or a few, $m^\mu$ macroscopic $(O(N^0))$ and the rest microscopic $(\leq 0(N^{-\frac{1}{2}}))$.  This enables us to obtain self-consistent equations for the overlaps with the nominated (potentially macroscopically overlapped or condensed) patterns
\begin{equation}
m^\mu=N^{-1}\sum_i\xi^\mu_i\langle\langle\sigma_i\rangle T\rangle_{\{\xi\}};=1,\cdots s,
\end{equation}
where the $1,...s$ label the nominated patterns, $\langle~\rangle_T$ denotes the thermal average at fixed $\{\xi\}$, and $\langle~\rangle_{\{\xi\}}$ denotes an average over the other (uncondensed) patterns, together with a spin-glass like order parameter
\begin{equation}
q=\langle N^{-1}\sum_i\langle\sigma_i\rangle^2_T\rangle_{\{\xi\}}
\end{equation}
and a mean-square average of the overlaps with the un-nominated patterns.  Retrieval corresponds to a solution with just one $m^\mu$ non-zero.

Explicitly, averaging over random patterns yields
\begin{eqnarray}
\langle Z^n\rangle &=& \exp (-np\beta/2)\sum_{\{\sigma^{\alpha}\}}\int\Pi^s_\mu\Pi^n_\alpha\{dm^{\mu\alpha}(\beta N/2\pi)^{\frac{1}{2}}\exp[-N(\beta\sum_\alpha(m^{\mu\alpha})^2/2 \nonumber \\ 
&~ ~ & +N^{-1}\sum_i\ln \cosh (\beta \sum_\alpha m^{\mu\alpha}\sigma^\alpha_i))]\}. \label{eq:Zn}
\end{eqnarray}

To proceed further we separate out the condensed and non-condensed patterns.  For the non-condensed patterns, $\mu > s$, only small $m^\mu$ contribute and the corresponding $\ln \cosh$ can be expanded to second order to approximate.
\begin{equation}
\Pi_{\mu>s}\exp[-\sum_i\ln\cosh(\beta\sum_\alpha m^{\mu\alpha}\sigma^\alpha_i)] \to \Pi_{\mu>s}\exp[\frac{\beta^2}{2}\sum_{\alpha\beta}m^{\mu\alpha}m^{\mu\beta}\sum_i\sigma^\alpha_i\sigma^\beta_i].\label{eq:A1}
\end{equation}
$\sum\sigma^\alpha_i\sigma^\beta_i$ may be effectively decoupled by the introduction of a spin-glass like order parameter $q^{\alpha\beta}$ via the identities
\begin{eqnarray}
1&=&\int dq^{\alpha\beta}\delta(q^{\alpha\beta}-N^{-1}\sum_i\sigma^\alpha_i\sigma^\beta_i) \nonumber \\
&=& \int dq^{\alpha\beta}\int\frac{d{\hat{q}}}{2\pi}^{\alpha\beta}\exp(i{\hat{q}}^{\alpha\beta}(q^{\alpha\beta}-N^{-1}\sum_i\sigma^\alpha_i\sigma^\beta_i)),
\end{eqnarray}
whence (\ref{eq:A1}) becomes
\begin{eqnarray}
&&\int\Pi_{(\alpha\beta)}dq^{\alpha\beta}\frac{d{\hat{q}}}{2\pi}^{\alpha\beta}\exp(i\sum_{(\alpha\beta)}{\hat{q}}^{\alpha\beta}(q^{\alpha\beta}-N^{-1}\sum_i\sigma^\alpha_i\sigma^\beta_i) \nonumber \\
&+& N\beta^2\sum_{u>s}\sum_{(\alpha\beta)}q^{\alpha\beta}m^{\mu\alpha}m^{\mu\beta} 
+\frac{N\beta^2}{2}\sum_{u>s}\sum_\alpha(m^{\mu\alpha})^2).
\end{eqnarray}

In (\ref{eq:Zn}) the $m^{\mu\alpha}; \mu>s$ integrations now yield the $\sigma$-independent result $(2\pi/N\beta)^{\frac{1}{2}(p-s}({\rm{det}}~A)^{-\frac{1}{2}(p-s)}$, where
\begin{equation}
A^{\alpha\beta}=(1-\beta)\delta_{\alpha\beta}-\beta q^{\alpha\beta},
\end{equation}
while the $\sigma_i$ contributions enter in the separable form
\begin{equation}
\exp[\sum_i(\sum^s_{\mu=1}\ln\cosh(\beta\sum_\alpha m^{\mu\alpha}\sigma^\alpha_i) - iN^{-1}\sum_{(\alpha\beta)}{\hat{q}}^{\alpha\beta}\sigma^\alpha_i\sigma^\beta_i)].
\end{equation}
Further anticipating the result that the relevant ${\hat{q}}$ scales as $p=\alpha N$ has the consequence that re-scaling ${\hat{q}}=i\beta^2 p r$
\begin{equation}
\langle Z^n\rangle_{\{\xi\}} = (\beta N/2\pi)^{n/2}\int \Pi^n_{\mu,\alpha=1}dm^{\mu\alpha}\int\Pi_{(\alpha\beta)}dq^{\alpha\beta}dr^{\alpha\beta}e^{-N\beta\Phi}\label{eq:Zna}
\end{equation}
where $\Phi$ is intensive given by
\begin{eqnarray}
&&\Phi(\{m^\alpha\},\{q^{\alpha\beta}\},\{r^{\alpha\beta}\}) \nonumber \\
&=&\frac{np}{2N}+\frac{1}{2}\sum_\alpha\sum^s_{\mu=1}(m^{\mu\alpha})^2+\frac{(p-s)}{2\beta N}{\rm{Tr}}\ln A+\alpha\beta\sum_{(\alpha\beta)}r^{\alpha\beta}q^{\alpha\beta} \nonumber \\
&-&\beta^{-1}\ln\sum_{\{\sigma^{\alpha}\}}\exp\{\sum^s_{\mu=1}\ln\cosh(\beta\sum m^{\mu\alpha}\sigma^\alpha)+\alpha\beta\sum_{(\alpha\beta)}r^{\alpha\beta}\sigma^\alpha\sigma^\beta\}.
\end{eqnarray}
(\ref{eq:Zna}) is thus extremally dominated.  At the extremum
\begin{eqnarray}
&&m^{\mu\alpha}=N^{-1}\sum_i\langle\xi^\mu_i\sigma^\alpha_i\rangle_\Phi ~ ~ ~ ~ ~ ~  ~ ~ ~ ~  ;\mu=1,...s \\
&&q^{\alpha\beta}=N^{-1}\sum_i\langle\sigma^\alpha_i\sigma^\beta_i\rangle_\Phi ~ ~ ~ ~ ~  ~ ~ ~ ~ ~ ~ ;\alpha \not= \beta \\
&&r^{\alpha\beta}=p^{-1}\sum_{\mu>1}\sum_i\langle m^{\mu\alpha}_im^{\mu\beta}_i\rangle_\Phi ~ ~ ;\alpha\not=\beta.
\end{eqnarray}
Within a replica-symmetric ansatz $m^{\mu\alpha}=m^\mu,q^{\alpha\beta}=q,r^{\alpha\beta}=r$, self-consistency equations follow relatively straightforwardly.  For the retrieval situation in which only one $m^\mu$ is macroscopic (and denoted by $m$ below) they are
\begin{eqnarray}
&&m=\int\frac{dz}{\sqrt{2\pi}}\exp(-z^2/2)\tanh[\beta(z\sqrt{\alpha r}+m)] \\
&&q=\int\frac{dz}{\sqrt{2\pi}}\exp(-z^2/2)\tanh^2[\beta(z\sqrt{\alpha r}+m)] \\
&&r=q(1-\beta(1-q))^{-2}
\end{eqnarray}

Retrieval corresponds to a solution $m\not=0$.  There are two types of non-retrieval solution, (i) $m=0, q=0$, called paramagnetic, in which the system samples all of phase space, (ii) $m=0,q\not=0$, the spin glass solution, in which the accessible phase space is restricted but not correlated with a pattern.  Fig. 10 shows the phase diagram \cite {AGS}; retrieval is only possible provided the combination of stochastic noise $T$ and pattern interference noise $\alpha$ is not too great.  There are also spurious solutions with more than one $m^\mu\not=0$, but these are not displayed in the figure.

As in the case of the SK spin glass one can check for instability against replica symmetry breaking.  RS turns out to be unstable in the spin glass region and in a small part of the retrieval region of $(T,\alpha)$ space near the maximum $\alpha$ for retrieval.  However RSB gives rise to only relatively small changes in the critical retrieval capacity.

\begin{figure}[tb!]
\centerline{\piccie{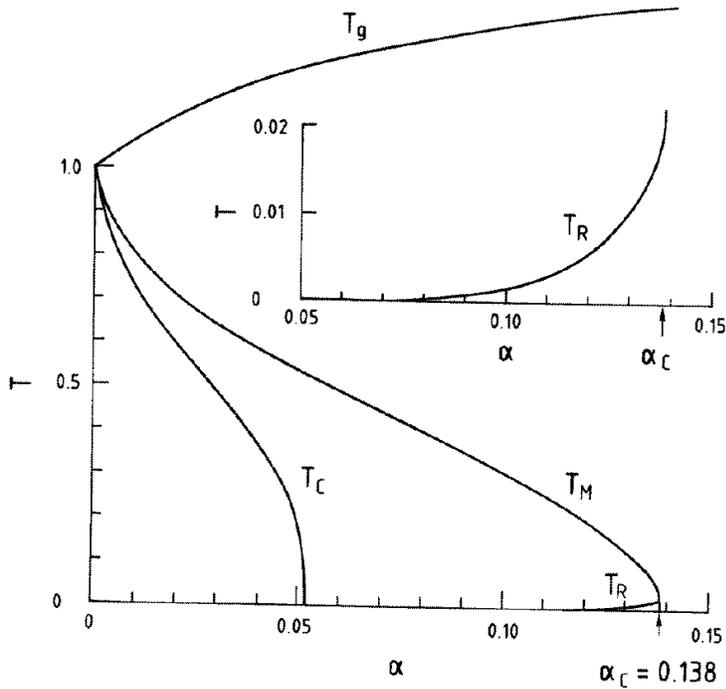}{3.6in}}
\caption{Phase diagram of the Hopfield model (after Amit et. al $^{36})$.  $T_m$ indicates the limit of retrieval solutions, between $T_m$ and $T_g$ there are spin-glass like non-retrieval solutions, above $T_g$ only paramagnetic non-retrieval.  $T_c$ denotes a thermodynamic transition, at which retrieval states become lowest in free energy, rather than just attractors.  $T_R$ shows the limit of replica-symmetry breaking in the retrieval phase.}
\end{figure}

\subsection{Statistical Mechanics of Learning}

In the last section we considered the problem of assessing the retrieval capability of a system of given architecture, local update rule and algorithm for $\{J_{ij}\}$.  Another important issue is the converse; how to choose/train the $\{J_{ij}\}$, in order to achieve the best performance.  Various such performance measures are possible; for example, in a recurrent network one might ask for the best overlap improvement in one sweep, or the best asymptotic retrieval, or the largest size of attractor basin, or the largest storage capacity, or the best resistance to damage; in a feedforward network trying to learn a rule from examples one might ask for the best performance on the examples presented, or the best ability to generalize.  Statistical mechanics as developed for spin glasses has played an important role in assessing what is achievable in such optimization and also provides a possible mechanism for achieving such optima (although there may be other algorithms which are quicker to attain the goals which have been shown to be accessible).

Thus in this section we discuss the statistical physics of optimization, as applied to neural networks.  Similar techniques apply to other optimization problems.\vskip0.10in

\noindent
{\it{Statistical physics of optimization}}

Consider a problem specifiable as the minimization of a function $E_{\{{\bm{a}}\}}(\{{\bm{b}}\})$ where the $\{{\bm{a}}\}$ are quenched parameters and the $\{{\bm{b}}\}$ are the variables to be adjusted, and furthermore, the number of possible values of $\{{\bm{b}}\}$ is very large.  In general such a problem is hard.  One cannot try all combinations of $\{{\bm{b}}\}$ since there are too many.  Nor can one generally find a successful iterative improvement scheme in which one chooses an initial value of $\{{\bm{b}}\}$ and gradually adjusts the value so as to accept only moves reducing $E$.  Rather, if the set $\{{\bm{a}}\}$ imposes conflicts, the system is likely to have a `landscape' structure for $E$ as a function of $\{{\bm{b}}\}$ which has many valleys ringed by ridges, so that a downhill start from most starting points is likely to lead one to a secondary higher-$E$ local minimum and not a true global minimum, or even a close approximation to it.

To deal with such problems computationally the technique of simulated annealing was invented \cite{KGV}.  It simulated the thermal excitation procedure used by a metallurgist to anneal out the defects which typically result from rapid quenches.  Specifically, one treats $E$ as a microscopic `energy', invents a complementary annealing temperature $T_A$, and simulates a stochastic thermal dynamics in $\{{\bm{b}}\}$ which, in principle, iterates to a distribution of the Gibbs form
\begin{equation}
p(\{{\bm{b}}\}) \sim \exp (-E(\{{\bm{b}}\})/T_A),\label{eq:Gibbs}
\end{equation} 
Then one reduces $T_A$ gradually to zero, to try to achieve the ground state.

The actual dynamics has some freedom - for example, for discrete variables Monte Carlo simulations with either a heat bath or a Metropolis algorithm both lead to (\ref{eq:Gibbs}).  For continuous variables Langevin dynamics with $\frac{\partial{\bm{b}}}{\partial t} = - \nabla_{\bm{b}}E({\bm{b}})+\eta(t)$, where $\eta(t)$ is white noise of strength $T_A$, would also be appropriate.  Of course, for a frustrated and disordered $E(\{{\bm{b}}\})$ the slow relaxation and non-equilibration features discussed earlier make it difficult to achieve the true minimum in practice.

Computational simulated annealing is used to determine specific $\{{\bm{b}}\}$ given specific $\{{\bm{a}}\}$.  It is also of interest, however, to consider generically what is achievable, averaged over all equivalently chosen $\{{\bm{a}}\}$.  

Hence we turn to the analytic equivalent of simulated annealing.  We define a generalized partition function
\begin{equation}
Z_A=\sum_{\{{\bm{b}}\}} \exp (-E(\{{\bm{b}}\})/T_A),
\end{equation}
from which the average `energy' at temperature $T_A$ follows from
\begin{equation}
\langle E\rangle_{T_{A}}=-\frac{\partial}{\partial\beta_A}\ln Z_A; ~ ~ ~ \beta_A=T_A^{-1}
\end{equation}
and the minimum $E$ from the zero `temperature' limit, 
\begin{equation}
E_{{\rm{min}}}=\lim_{T_{A}\to 0}\langle E\rangle_{T_{A}}.
\end{equation}

As noted earlier, we are often interested in typical behaviour, as characterized by averaging the result over a random choice of $\{{\bm{a}}\}$ from some characterizing distribution.  Hence we want $\langle\ln Z_A\rangle_{\{{\bm{a}}\}}$, which naturally suggests the use of replicas again.

In fact, the replica procedure has been used to study several hard combinatorial optimization problems, such as various graph partitioning \cite{FA,KS2,WS} and travelling salesman problems.  Here, however, we shall concentrate on neural network applications.\vskip0.10in

\noindent
{\it {Cost functions dependent on stability fields}}

One important class of training problems for pattern-recognition neural networks is that in which the objective can be defined as minimization of a cost function dependent on patterns and synapses only through the stability fields; that is, in which the `energy' to be minimized can be expressed in the form
\begin{equation}
E^A_{\{\xi\}}(\{J\})=-\sum_\mu\sum_i g(\Lambda^\mu_i);~ ~ ~ ~ ~ ~\Lambda^\mu_i=\xi^\mu_i\sum_{j\not=i}J_{ij}\xi^\mu_j/(\sum_{j\not=i}J^2_{ij})^{\frac{1}{2}}.\label{eq:stabcost}
\end{equation}

Before discussing general procedure, some examples of $g(\Lambda)$ might be in order.  

The original application of this technique to neural networks concerned the maximum capacity for stable storage of patterns in a network satisfying the local update rule \cite{G1}
\begin{equation}
\sigma^\prime_i = ~{\rm{sgn}}~ (\sum_{j\not=i}J_{ij}\sigma_j).
\end{equation}
Stability is determined by the $\Lambda^\mu_i$; if $\Lambda^\mu_i>0$, the input of the correct bits of pattern $\mu$ to site $i$ yields the correct bit as output.  Thus a pattern $\mu$ is stable under the network dynamics if $\Lambda^\mu_i>0; {\rm{all}}~i$.  A possible performance measure is therefore given by (\ref{eq:stabcost})
\begin{equation}
g(\Lambda)=-\Theta(-\Lambda). \label{eq:minstab}
\end{equation}
$g(\Lambda^\mu_i)$ is thus non-zero (and negative) when pattern $\mu$ is not stably stored at site $i$.  Choosing the $\{J_{ij}\}$ such that the minimum $E$ is zero ensures stability.  The maximum capacity for stable storage is the limiting value of $p/N=\alpha$ for which stable storage is possible.

An extension is to maximal stability \cite{GD}.  In this case the performance measure employed is $g(\Lambda)=-\Theta(\kappa-\Lambda)$ and the search is for the maximum value of $\kappa$ for which $E_{min}$ can be held to zero for any capacity $\alpha$, or, equivalently, the maximum capacity for which $E_{min}=0$ for any $\kappa$.

Yet another example is to consider a system trained to give the greatest increase in overlap with a pattern in one step of the dynamics, when started in a state with overlap $m_t$  For the update rule $\sigma_i\to\sigma^\prime_i=~{\rm{sgn}}~(h_i+Tz)$ where $z$ is stochastically Gaussian-random, the appropriate performance function, averaged over all specific starting states of overlap $m_t$, is \cite {WS2}
\begin{equation}
g(\Lambda)={\rm{erf}}\left\{\frac{(m_t\Lambda)}{\sqrt{2(1+m^2_t+T^2)}}\right\}.
\end{equation}\vskip0.10in

\noindent
{\it{Methodology}}

Let us now turn explicitly to the analytic minimization of a cost function of the general form
\begin{equation}
E^A_{\{\xi\}}(\{J\})=-\sum_\mu g(\Lambda^\mu); ~ ~ ~ ~ ~ \Lambda^\mu=\xi^\mu_0\sum^N_{j=1}J_j\xi^\mu_j,
\end{equation}
with respect to $J_j$ which satisfy spherical constraints $\sum J^2_j=N$.
The $\{\xi\}$ are random quenched $\pm 1$.  The result is expressed averaged over the choice of the $\{\xi\}$.  Thus we require $\langle\ln Z_{\{\xi\}}\rangle_{\{\xi\}}$ where
\begin{equation}
Z_{\{\xi\}}=\int\Pi_jdJ\delta(\sum_jJ^2_j-N)\exp(\beta_A\sum_\mu g(\Lambda^\mu)).
\end{equation}
In order to evaluate the $\{\xi\}$ average we separate out the explicit $\xi$ dependence in $g(\Lambda)$ via delta functions $\delta(\lambda^\mu-\xi^\mu\sum_jJ_j\xi^\mu_j/N^{\frac{1}{2}})$ and express all the delta functions in exponential integral representation, 
\begin{eqnarray}
Z_{\{\xi\}}=\int\Pi_jdJ_j&&\int\frac{d\epsilon}{2\pi}\exp(i\epsilon(\sum_jJ^2_j-N))\Pi_\mu\int d\lambda^\mu\frac{d\phi^\mu}{2\pi}\exp(\beta_Ag(\lambda^\mu)) \nonumber \\
&&\exp(i\phi^\mu(\lambda^\mu - \xi^\mu\sum_jJ_j\xi^\mu_j/N^{\frac{1}{2}})).
\end{eqnarray}

Replica theory requires $\langle Z^n_{\{\xi\}}\rangle_{\{\xi\}}$ and therefore the introduction of a dummy replica index on each of the  $J,\epsilon,\lambda$ and $\phi$; we use $\alpha=1,...n$.  For the case in which all the $\xi$ are independently distributed and equally likely to be $\pm 1$, the $\xi$ average involves
\begin{equation}
\langle\exp(-i\phi^{\mu\alpha}\xi^\mu\sum_jJ^\alpha_j\xi^\mu_j/N^{\frac{1}{2}})\rangle_{\{\xi\mu\}}=\cos(\sum_\alpha\phi^{\mu\alpha} J^\alpha_jN^{-\frac{1}{2}}).
\end{equation}
For large $N$ (and $\phi J\leq O(1))$ the cosine can be approximated (after expansion and re-exponentiation) by 
\begin{equation}
\cos(\sum_\alpha\phi^{\mu\alpha} J^\alpha_jN^{-\frac{1}{2}})=\exp(-(\sum_\alpha\phi^{\mu\alpha}J^\mu_j)^2/2N).
\end{equation}
Thus
\begin{equation}
\langle Z^n\rangle_{\{\xi\}}=\int\Pi_\alpha\frac{d\epsilon^\alpha}{2\pi} \Pi_jdJ^\alpha_j\Pi_\mu d\lambda^{\mu\alpha}\frac{d\phi}{2\pi}^{\mu\alpha}\exp\Phi(\{\xi^\alpha\},\{J^\alpha_j\},\{\lambda^{\mu\alpha}\},\{\phi^{\mu\alpha}\})
\end{equation}
where 
\begin{eqnarray}
\Phi &=& i\sum_\alpha\epsilon^\alpha(\sum_j(J^\alpha_j)^2-N)+i\sum_{\mu,\alpha}\phi^{\mu\alpha} \lambda^{\mu\alpha}-\frac{1}{2}\sum_{\mu\alpha}(\phi^{\mu\alpha})^2 \nonumber \\
&-& \sum_\mu\sum_{\alpha<\beta}\phi^\alpha_\mu\phi^\beta_\mu(N^{-1}\sum_jJ^\alpha_jJ^\beta_j)+\beta_A\sum_{\mu\alpha}g(\lambda^{\mu\alpha}).
\end{eqnarray}
We eliminate the term in $\sum_jJ^\alpha_jJ^\beta_j$ in favour of a spin glass like order parameter $q^{\alpha\beta}$
\begin{eqnarray}
1&=&\int dq^{\alpha\beta}\delta(q^{\alpha\beta}-N^{-1}\sum_jJ^\alpha_jJ^\beta_j) \\
&=&(N/2\pi)\int dx^{\alpha\beta}dq^{\alpha\beta}\exp(ix^{\alpha\beta}(Nq^{\alpha\beta}-\sum_jJ^\alpha_jJ^\beta_j)).
\end{eqnarray}
The $j=1,...N$ and $\mu = 1,...p$ contribute only multiplicatively and 
\begin{eqnarray}
\langle Z^n\rangle_{\{\xi\}}=\int\Pi_{\alpha<\beta}\frac{dx^{\alpha\beta}}{2\pi}dq^{\alpha\beta}\Pi_\alpha\frac{d\epsilon^\alpha}{2\pi}\exp[N(&-&\sum_{\alpha<\beta}q^{\alpha\beta}x^{\alpha\beta} + G_J(\{\epsilon^\alpha\},\{x^{\alpha\beta}\}) \nonumber \\
&+&(p/N)G_\xi(\{q^{\alpha\beta}\}))]; \nonumber \label{eq:B13}
\end{eqnarray}
\begin{equation}
\exp G_J(\{\epsilon^\alpha\},\{x^{\alpha\beta}\})=\int\Pi_\alpha dJ^\alpha\exp(-\sum_\alpha\epsilon^\alpha((J^\alpha)^2-1)+\sum_{\alpha<\beta}x^{\alpha\beta}J^\alpha J^\beta),
\end{equation}
\begin{equation}
\exp G_\xi(\{q^{\alpha\beta}\})=\int\Pi_\alpha d\lambda^\alpha\frac{d\phi}{2\pi}^\alpha\exp(\sum_\alpha[\beta_Ag(\lambda^\alpha)+i\lambda^\alpha\phi^\alpha-(\phi^\alpha)^2/2-\sum_{\alpha<\beta}q^{\alpha\beta}\phi^\alpha\phi^\beta).
\end{equation}
Since $g_J$ and $G_\xi$ are intensive, as is $p/N=\alpha$ for the situation of interest, (\ref{eq:B13}) is dominated by the maximum of its integrand.

In the replica symmetric ansatz $\epsilon^\alpha=\epsilon, x^{\alpha\beta}=x$ and $q^{\alpha\beta}=q$.  In the limit $n\to 0$, elimination of $\epsilon$ and $x$ at the saddle point $\partial\Phi/\partial\epsilon = \partial\Phi/\partial x = \partial\Phi/\partial q=0$ yields
\begin{eqnarray}
&&\langle\ln Z\rangle_{\{\xi\}} = N ~{\rm{ext}}_q(\frac{1}{2}\ln [2\pi(1-q)]+(2(1-q))^{-1} \nonumber \\
&&+\alpha\int Dt\ln(2\pi(1-q))^{-\frac{1}{2}}\int d\lambda\exp[\beta_Ag(\lambda)-(\lambda-\sqrt{qt})^2/2(1-q)]) ~ ~ \\
&&{\rm{where}}~ Dt=dt\exp(-t^2/2)/\sqrt{2\pi}.
\end{eqnarray}

In the low temperature limit, $\beta_A\to\infty, ~ ~ q\to 1$ and $\beta_A(1-q)\to\gamma$, independent of $T_A$ to leading order.  The integration over $\lambda$ can then be simplified by steepest descent, so that
\begin{equation}
\int d\lambda\exp(\beta_Ag(\lambda)-(\lambda-t)^2/2(1-q))\to\exp(\beta_A(g(\tilde{\lambda}(t))-(\tilde{\lambda}(t)-t)^2/2\gamma))
\end{equation}
where $\tilde{\lambda}(t)$ is the value of $\lambda$ which maximizes $(g(\lambda)-(\lambda-t)^2/2\gamma)$; i.e. the inverse function of $t(\tilde{\lambda})=\tilde{\lambda}-\gamma g^\prime(\lambda)|_{\lambda=\tilde{\lambda}}.$
Extremizing with respect to $q$ gives the (implicit) determining equation for $\gamma$
\begin{equation}
\int Dt(\tilde{\lambda}(t)-t)^2=\alpha
\end{equation}

The average minimum cost follows from 
\begin{equation}
\langle E_{min}\rangle_{\{\xi\}}=\lim_{\beta_{A}\to\infty}\frac{\partial}{\partial\beta_A}\langle\ln Z\rangle_{\{\xi\}}.
\end{equation}
Similarly, any measure $\langle\langle\sum^p_{\mu=1}f(\Lambda^\mu)\rangle_{T_{A}}\rangle_{\{\xi\}}$ may be obtained from $<\ln Z>$ by means of generating the functional procedure.  Alternatively, they follow from the local field distribution $p(\lambda)$ defined by 
\begin{equation}
\rho(\lambda)=\langle\langle p^{-1}\sum^p_{\mu=1}\delta(\lambda - \Lambda^\mu)\rangle_{T_{A}}\rangle_{\{\xi\}}
\end{equation}
which is given by
\begin{equation}
\rho(\lambda)=\int Dt\delta(\lambda-\tilde{\lambda}(t)).
\end{equation}

Just as in the replica analysis of retrieval, the assumption of replica symmetry for $q^{\alpha\beta}$  needs to be checked and a more subtle ansatz employed when it is unstable against small fluctuations $q^{\alpha\beta}\to q+\eta^{\alpha\beta};\eta^{\alpha\beta}$ small.  In fact, it should also be tested even when small fluctuations are stable (since large ones may not be).  Such effects, however, seem to be absent or small for many cases of continuous $\{J_{ij}\}$, while for discrete $\{J_{ij}\}$ they are more important\footnote{for discrete $\{J_{ij}\}$ there is first order replica-symmetry breaking$^{45}$ at a temperature higher than that at which there is instability against small fluctuations.}.\vskip0.10in

\noindent
{\it{Learning a rule}}

So far, our discussion of optimal learning has concentrated on recurrent networks and on training perceptron units for association of given patterns.  Another important area of practical employment of neural networks is as expert systems, trained to try to give correct decisions on the basis of many observed pieces of input data.  More precisely, one tries to train a network to reproduce the results of some (usually-unknown) rule relating many-variable inputs to few-variable outputs, on the basis of training with a few examples of input-output sets arising from the operation of the rule (possibly with error in this training data).

To assess the potential of an artifical network of some structure to reproduce the output of a rule on the basis of examples, one needs to consider the training of the network with examples of input-output sets generated by known rules, but without the student network receiving any further information, except perhaps the probability that the teacher rule makes an error (if it is allowed to do so).

Thus let us consider first a deterministic teacher rule $\eta=V(\{\xi\}),$ relating $N$ elements of input data $(\xi^\mu_1,\xi^\mu_2...\xi^\mu_N)$ to a single output $\eta^\mu$, being learned by a deterministic student network $\eta=B(\{\xi\})$. $B$ is known whereas $V$ is not.  Training consists of modifying $B$ on the basis of examples drawn from the operation of $V$.  Problems of interest are to train $B$ to give (i) the best possible performance on the example set, (ii) the best possible performance on any random sample drawn from the operation of $V$, irrespective of whether it is a member of the training set or not.  The first of these refers to the ability of the student to learn what he is taught, the second to his ability to generalise from that training.

The performance on the training set $\mu=1,...p$ can be assessed by a training error 
\begin{equation}
E_t=\sum^p_{\mu=1}e(B(\{\xi^\mu\}),V(\{\xi^\mu\}))
\end{equation}
where $e(x,y)$ is zero if $x=y$, positive otherwise.

A common choice for $e$ is quadratic in the difference $(x-y)$.  With the scaling $e(x,y)=(x-y)^2/4$ one has for binary outputs, $\eta=\pm1$, $e(x,y)=\Theta(-xy)$, so that if $B(\{\xi\})$ is a perceptron, 
\begin{equation}
B(\{\xi\})=~{\rm{sgn}}~(\sum_jJ_j\xi_j),
\end{equation}
then $e^\mu=\Theta(-\Lambda^\mu)$ where now $\Lambda^\mu=\eta^\mu\sum_j J_j\xi^\mu_j/(\sum_jJ^2_j)^{\frac{1}{2}}$, making $e_t$ analgous to $E^A$ of (\ref{eq:stabcost}) with `performance function' (\ref{eq:minstab}).  This we refer to as minimal stability learning.  Similarly, one can extend the error definition to $e^\mu=\Theta(\kappa-\Lambda^\mu)$ and, for learnable rules, look for the solution with the maximum $\kappa$ for zero training error.  This is maximal stability learning.

Minimizing $e_t$ can proceed as discussed above, either simulationally or analytically.  Note, however, that for the analytic study of average performance the $(\eta,\xi)$ combinations are now related by the rule $V$, rather than being completely independent.  The generalization error $\epsilon_g=\langle e(B(\{\xi\}),V(\{\xi\})\rangle _\delta$ follows from the resultant distribution $\rho(\Lambda)$.

\section{Dynamical replica theory}

Let us now turn to a different approach to macroscopic dynamics, namely an attempt to find good approximate (or exact) closed autonomous flow equations for macroscopic observables, starting from instantaneous stochastic microdynamics \cite{CS,CS2,LCS}.

Specifically, let us concentrate on a system with Ising variables obeying random sequential Glauber dynamics as in 4.1 (i).  With appropriate coarse-graining this leads to the master equation
\begin{equation}
\frac{d}{dt}p_{t}({\bsigma})=\sum_{k=1}^{N} \left[p_{t}(F_{k}\bsigma)W_{k}(F_{k}\bsigma) - p_{t}(\bsigma)W_{k}(\bsigma)\right]
\label{eq:master}
\ee
where $F_k$ is the spin-flip operator 
\begin {equation} 
F_k\Phi(\bsigma) = \Phi (\sigma_i,...,-\sigma_k,...,\sigma_N), 
\label{eq:spinflip}
\end{equation}
$W_k(\bsigma)$ is the transition rate 
\be
W_k(\bsigma) = {1\over 2}[1-\sigma_k\tanh(\beta h_k(\bsigma))],  
\label{eq:transition}
\ee
and we are again using the vector notation $\bsigma\ = (\sigma_1,...,\sigma_N)$.

From (\ref{eq:master}) we may derive an equation for the evolution of the macrovariable probability distribution 

\begin{equation}
P_t[\bomega]=\sum_{\bsigma} p_t (\bsigma) \delta[\bomega-\bomega(\bsigma)]; \ \ \  \bomega\equiv(\Omega_1,...\Omega_n)
\label{eq:Pomega}
\end{equation}
in the form
\begin{equation}
\frac{d}{dt} P_t[\bomega]=
\sum_{\ell\geq 1} \frac{(-1)}{\ell !}^\ell \sum^n_{k_{1}=1}.. \sum^n_{k_{\ell}=1} 
\frac{\partial^\ell}{\partial\Omega_{k_1}...\partial\Omega_{k_{\ell}}} P_t[\bomega]F^{(\ell)}_{k_{1}..k_{\ell}} [\bomega ;t]
\label{eq:dPomegadt}
\end{equation}
where
\begin{equation}
F^{(\ell)}_{k_{1}...k_{\ell}}[\bomega ;t]=\langle\sum^N_{j=1}W_j(\bsigma)\Delta_{jk_{1}}(\bsigma)...\Delta_{jk_{\ell}}(\bsigma)\rangle_{\bomega;t}
\ \ \ \ \ \Delta_{jk}(\bsigma)\equiv\Omega_k(F_j\bsigma)-\Omega_k(\bsigma)
\label{eq:F}
\end{equation}
and the notation $\langle\rangle_{\bf{\Omega};t}$ refers to a sub-shell average
\begin{equation}
\langle f(\bsigma)\rangle _{\bomega;t}\equiv \frac{\sum_{\bsigma}p_t({\bsigma})\delta[{\bomega}-{\bomega}({\bsigma})]f({\bsigma})}{\sum_{\bsigma}p_t({\bsigma})\delta[{\bomega}-{\bomega}({\bsigma)]}} \ \ .
\label{eq:f}
\end{equation}

In several cases of interest and for finite times, only the first term on the right hand side of (\ref{eq:dPomegadt}) survives in the limit $N\to\infty$, yielding the deterministic flow
\begin{equation}
\frac{d}{dt}{\bomega}_t = \langle\sum_iW_i({\bsigma})[{\bomega}(F_i{\bsigma}) - {\bomega}({\bsigma})]\rangle _{{\bomega};t} \ \ .
\label{eq:domegadt}
\end{equation}
In general this does not yet constitute a closed set of equations due to the appearance of $p_t(\bsigma)$ in the sub-shell average.  However, we may attempt to find an appropriate choice of $\bomega$ for which closure may be attained either exactly or approximately. Ideally we would like $\bomega$ to be as low-dimensional as possible.

One particularly simple example occurs for the case of an infinite range Ising ferromagnet; $J_{ij}={J_0/N}$. In this case the magnetization $m=N^{-1}\sum_{i}\sigma_{i}$ suffices alone as a macrovariable whose evolution is deterministic and closed,
\be
\frac{d}{dt}m=\tanh(\beta J_0 m)-m,
\label{eq:ferromagnet}
\ee
and yields the usual mean field solution in the steady state limit
\be
\frac{d}{dt}m=0~~~~~\rightarrow ~~~~~m=\tanh(\beta(J_0 m)).
\label{eq:mft}
\ee

A greater challenge is posed by problems with sufficient disorder and frustration, such as those given by 
\begin{equation}
J_{ij}=J_o/N+Jz_{ij}/\sqrt{N};\ \ \ \ \  \ \ \langle z_{ij}\rangle =0 \ \ \ \langle {z_{ij}}^2\rangle =1 ;\ \ \ \ i\neq j.
\label{eq:exchange}
\end{equation}
where $z_{ij}$ is a quenched random parameter. This is the case for two particular model problems of interest; the Sherrington-Kirkpatrick (SK) spin glass and the Hopfield neural network.  In the SK model the $\{z_{ij}\}$ are chosen randomly from a Gaussian distribution.
In the Hopfield model, concentrating for simplicity on the region of phase space within the basin of attraction of one pattern, taken arbitrarily as $\mu=1$, it is convenient to apply the gauge transformation $\sigma_i\to\sigma_i\xi_i^1, \ \  J_{ij}\to \xi^1_i\xi^1_j J_{ij},$ to re-write (\ref{eq:Hopfield}) in the form above with 
\begin{equation}
z_{ij}=\frac{1}{\sqrt{p}} \sum^p_{\mu>1} \xi^1_i\xi^\mu_i\xi^1_j\xi^\mu_j \ \ \ \  J_o=1 \ \ \ \ J=\sqrt{\alpha} \ \ \ \ \langle\xi_i\rangle=0 \ \ .
\label{eq;Hopfield}
\end{equation}
It is straightforward to show that $m=N^{-1}\sum_{i}\sigma_{i}$ is insufficient for a closed macroscopic evolution for finite $J$, although it does suffice if $J\rightarrow 0$ as $N \rightarrow \infty$, as is the case for a Hopfield model with only one condensed pattern, storing only a less than extensive number of patterns $(\lim _{N \rightarrow\infty}{\alpha}=0)$. But how many macrovariables does one need and what are they?

Before giving an answer to the last question which appears to be at least very close to the truth, let us consider an intermediate step which is useful illustratively. Although our analysis applies to finite times, it is instructive to ask first about the long time steady state,assuming equilibration. For problems with detailed balance in their dynamics one knows that in the limit as $t\rightarrow\infty$ before $N\rightarrow\infty$ the microstate distribution takes the Boltzmann form $p_\infty(\bsigma) \sim \exp (-\beta H)$. This is the case in the above examples which have $J_{ij}=J_{ji}$, yielding
the Hamiltonian
\begin{equation}
H/N= -N^{-1}\sum_{i<j} J_{ij}\sigma_i\sigma_j \  
= -\frac{1}{2}J_0m^2({\bsigma}) - Jr({\bsigma}) + 0(N^{-1})
\label{eq:Hamiltonian}
\end{equation}
where
\be
r({\bsigma}) = N^{-3/2} \sum_{i<j}\sigma_iz_{ij}\sigma_j.
\ee
Thus, as long as $r(\bsigma) \sim O(1)$ it cannot be ignored in the set of $\bomega$. It is straightforward to show that $r(\bsigma) \sim O(1)$ for both the SK spin glass and the Hopfield model at finite storage ratio $\alpha$. Thus we shall first discuss an attempt to find a non-equilibrium macrodynamics in terms of $m,r$, alone, and show that it provides a reasonable but imperfect description. We shall then go on to a more sophisticated theory in terms of a  generalized order function which provides a very good fit to the results of microscopic simulation.   

\subsection{The simple version of the theory: two order parameters}

In this section we choose the minimal form 

\begin{equation}
{\bomega}^s({\bsigma})\equiv(\Omega_1({\bsigma}), \Omega_2({\bsigma})) = (m({\bsigma}), r({\bsigma})) \ \ .
\label{eq:minomega}
\end{equation}
The resultant $P_t[{\bomega}^s]$ does indeed satisfy a Liouville equation in the thermodynamic limit, yielding the deterministic flow equations

\begin{equation}
\frac{dm}{dt}=\int dzD_{m,r;t}(z)\tanh \beta (J_om+Jz)-m
\label{eq:m}
\end{equation}
\begin{equation}
\frac{dr}{dt}=\int dzD_{m,r;t}(z) z \tanh \beta(J_om+Jz)-2r
\label{eq:r}
\end{equation}
where $D_{m,r;t}(z)$ is the sub-shell averaged distribution of the disorder contributions to the local fields
\begin{equation}
D_{m,r;t}(z)=\lim_{N\to\infty}\frac{\sum_{\bsigma}p_t({\bsigma})\delta(m-m({\bsigma}))\delta(r-r({\bsigma}))N^{-1}\sum_i\delta(z-z_i({\bsigma}))}
{\sum_{\bsigma}p_t({\bsigma})\delta(m-m({\bsigma}))\delta(r-r({\bsigma}))}
\label{eq:noise}
\end{equation}
\begin{equation}
h_i({\bsigma})=J_om({\bsigma}) + Jz_{i}({\bsigma}) + 0(N^{-1}) \ \ \ \ \ \ z_i({\bsigma})=N^{-1/2}\sum_jz_{ij}\sigma_j \ \ .
\label{eq:hz}
\end{equation} 

As yet, because of the $p_t({\bsigma})$ in (\ref{eq:noise}) , equations (\ref{eq:m}) and (\ref{eq:r})  are not closed except in the disorder-free case $J=0$.  To close the equations we introduce two simple Ans\"atze:
(i) we assume that the evolution of the macrostate $(m,r)$ is self-averaging with respect to the specific microscopic realization of the disorder $\{z_{ij}\}$,
(ii) as far as evaluating $D(z)$ is concerned we assume equipartitioning of the microstate probability $p_t({\bsigma})$ within each $(m,r)$ shell.
The first of these Ans\"atze is well borne out by computer simulations of the microscopic dynamics and permits averaging $D(z)$ over pattern choices.  The second, which is clearly true as $t\rightarrow\infty$ since $p_{\infty}(\bsigma)$ depends only on $m$ and $r$ but can only be judged {\it a posteriori} for general time, eliminates memory effects beyond their reflection in $m,r$ and removes explicit time-dependence from $D$. Together these Ans\"atze give
\begin{equation}
D_{m,r;t}(z)\to D_{m,r}(z) = \left\langle \frac{\sum_{\bsigma}\delta(m-m({\bsigma}))\delta(r-r({\bsigma}))N^{-1}\sum_i\delta(z-z_i({\bsigma}))}{\sum_{\bsigma}\delta(m-m({\bsigma}))\delta(r-r({\bsigma}))}\right\rangle_{\{z_{ij}\}}
\label{eq:D}
\end{equation}
where $\left\langle\cdots\right\rangle_{{z_{\{ij\}}}}$ indicates an average over the quenched randomness. This yields closure of (\ref{eq:m}) and (\ref{eq:r}) since $D_{m,r}(z)$ now depends only upon the instantaneous values of $m, r$ and no longer on other microscopic measures of history. 

The actual evaluation of $D_{m,r}(z)$ from (\ref{eq:D}) remains a non-trivial exercise, but one which is amenable to solution by replica theory (as developed for the investigation of local field distributions in spin glasses \cite{TTCSS}).  After several manipulations it can be expressed in the form

\begin{equation}
D_{m,r}(z)=\lim_{n\to 0}\int\prod_{i,j}\prod_{\alpha,\beta=1...n}dx^\alpha_idy_j^{\alpha\beta}\exp[-N\Phi(m,r,z;\{x_i^\alpha\},\{y_j^{\alpha\beta}\})]
\label{eq:Dext}
\end{equation}
where the number of indices $i,j$ is finite and $\Phi$ is $O(N^0)$.  Because the argument of the exponential scales as $N$, the integral can be evaluated by steepest descents.  

The extremization is complicated and in its complete form involves significant subtleties, including an extension of those devised by Parisi for the analysis of the spin glass problem. It is discussed in detail elsewhere  \cite{CS,CS2}; here we note only a few salient results. Important among them is that in the steady state limit of $dm/dt = dr/dt =0$ the analysis yields the full thermodynamic results obtained from equilibrium analysis, including replica-symmetry breaking. For more general times explicit analysis to date has only been completed within the further Ansatz of replica-symmetry in the dynamic analogue of the spin glass order parameter $q^{\alpha\beta}$ which enters into the evaluation of  $D_{m,r}(z)$, but including a determination of the limit of its applicability against small replica-symmetry breaking fluctuations. 

The full analytic results for this case can be found in \cite{CS1,CS2}. Here we simply exhibit graphically the comparisons between theory and simulation for the Hopfield model for $\alpha=0.1$ and deterministic microdynamics. Fig 11 shows flows in $(m,r)$, with time implicit, and it may be observed that the comparison is quite good (but not perfect); it also clearly shows the need for (at least) two order parameters. On the other hand, Fig 12, which shows the dependence of $m$ and $r$ on $t$, demonstrates that the theory misses a slowing-down effect seen in the simulations for non-retrieving situations (ie. ones in which $\lim_{t\rightarrow\infty}m(t)=0)$. One may further note that the slowing-down occurs before the system crosses the limit of replica-symmetry stability against small fluctuations, suggesting that its origin lies elsewhere than in the breakdown of the RS Ansatz used in the evaluation of $D(z)$. Rather, one is driven to conclude that the problem lies in the loss of memory information inherent in the assumption of equipartitioning in the form used above. This implies that the set of $\bomega$ must be expanded beyond just $m$ and $r$, to include more microscopic effects.

\begin{figure}[tb!]
\centerline{\piccie{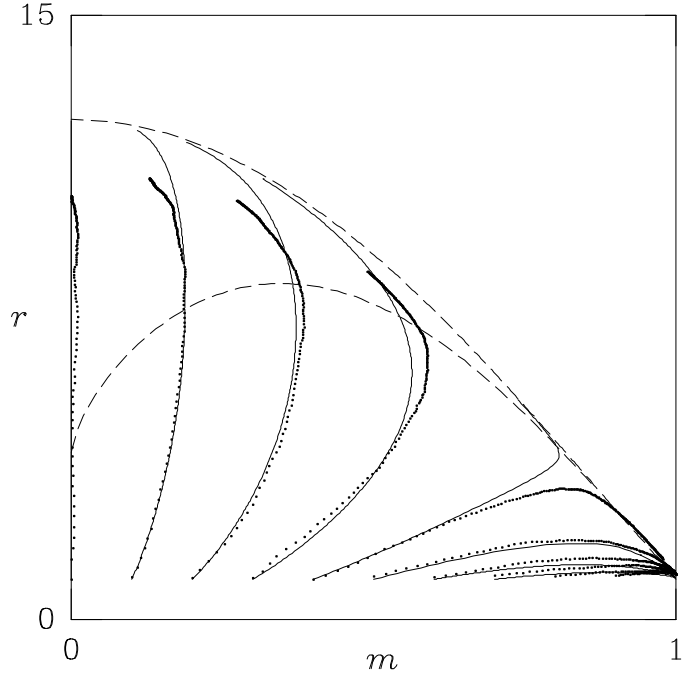}{3.5in}}
\caption{Macroscopic flow trajectories for a Hopfield model with
storage capacity $\alpha=0.1$ and deterministic microscopic dynamics
$(\beta = \infty)$; dots indicate simulations $(N=32000)$, solid lines
indicate analytic RS theory.  The outer dashed line is the boundary
predicted by RS theory; the inner dashed line indicates the onset of
instability against RS-breaking fluctuations, with stability on the
side closer to the origin (from $^{52}$).} 
\end{figure}
 
\subsection{The sophisticated version of the theory: order function dynamics}

To improve on the theory as developed in the last section requires broadening the range of order parameters. Addition of a finite number of extra observables is not however expected to give more than just minor improvement; rather, a qualitative change of philosophy would seem to be required. To this end we propose instead for $\bomega$ the joint distribution of the spins and the fields modifying them,  
\begin{equation}
{\cal{D}}(\s,h;\bsigma) 
= \frac{1}{N} \sum_i \delta_{\s,\sigma_i}
\delta\left[h \minus h_i (\bsigma)\right]. 
\label{eq:distribution}
\end{equation}
    
\begin{figure}[tb!]
\centerline{\piccie{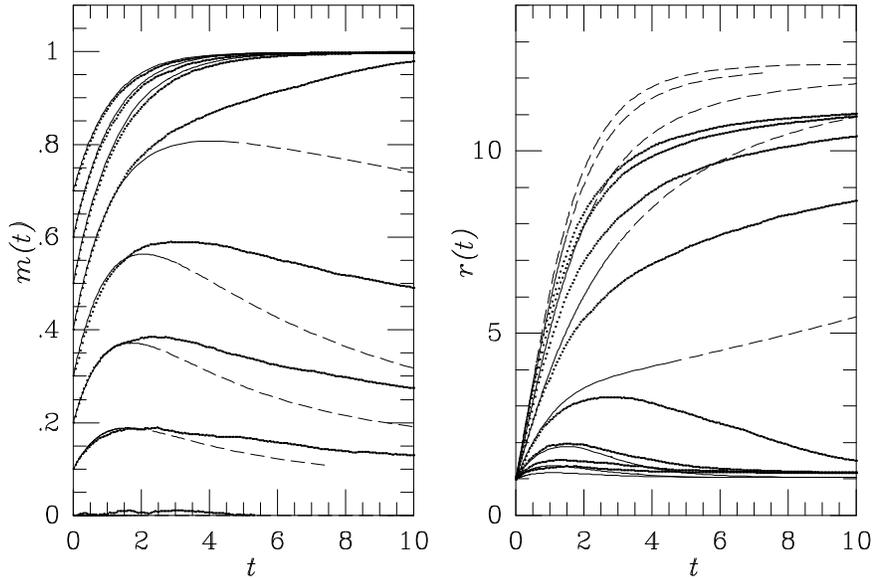}{3in}}
\caption{Temporal dependence of the order parameters for a Hopfield
model with storage $\alpha=0.1$ and zero-temperature dynamics; dots
indicate simulations $(N=32000)$, the other lines indicate RS theory
shown with solid lines where stable, dashed lines where unstable.
Time is measured in Monte Carlo steps per spin (from  $^{52}$).} 
\end{figure}

The equation of motion for ${\cal{D}}$ can be obtained by a (more complicated) analogue of that used in the last section to study the evolution of $m,r$.  Again this involves the use of self-averaging and equipartitioning to close the equations, but now the latter is much less restrictive (only microstates with the same distribution ${\cal{D}}$ are taken to contribute equally).  Details are given in \cite{LCS}.  Here, in Fig.13, we show only the quality of the result used to calculate the time-dependence of the binding energy of the SK model; agreement with computer simulations is excellent.

A detailed comparison of the two methods of allowing for macroscopic memory, the method of section 4 involving few two-time functions and the present method involving many single-time quantities \footnote{Note that ${\cal{D}}$ is a continuous function (i.e. effectively continuously many order parameters, one for each value of the function).}, remains incomplete.

\begin{figure}[tb!]
\centerline{\piccie{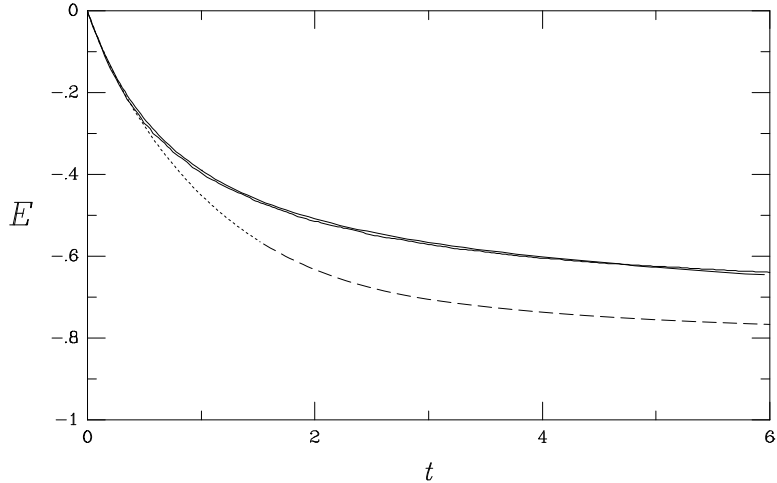}{2.5in}}
\caption{Evolution of the binding energy of the
Sherrington-Kirkpatrick spin glass $(J_o=0)$ from a random microscopic
start.  Comparison of simulations $(N=8000$, solid line) and
predictions of the simple two-parameter $(m,r)$ theory (RS stable,
dotted; RS unstable, dashed) and of the advanced order-function theory
of section V, for $\beta = \infty$.  Note that the two solid lines are
almost coincident (from  $^{53}$).}
\end{figure}

\section{Conclusion}

In these lectures I have tried to illustrate both the fundamental and the applicable character and interest of theoretical techniques developed to understand and quantify spin glasses.  Discussion has been restricted to mean field theory or infinite-ranged systems, because of space restrictions.  For further recent considerations of short-range spin glasses and other complementary experimental, simulational and theoretical studies the reader is referred to \cite{Young}.


\section*{References}
\begin{thebibliography}{99}

\bibitem{hist} For further historical and experimental background the reader is referred to\\
B.R. Coles, 1984, in ``Multicritical Phenomena'', eds. R. Pynn and A. Skjeltorp (Plenum Press), p363\\
J.A. Mydosh, 1993, ``Spin Glasses; an experimental introduction'' (Taylor and Francis)

\bibitem{CM}V. Canella and J.A. Mydosh, 1972, Phys. Rev. {\bf B6}, 4220

\bibitem{EA} S.F. Edwards and P.W. Anderson, 1975, J. Phys. {\bf F5}, 965

\bibitem{TT} J.L. Tholence and R. Tournier, 1974, J. Phys. (Paris) {\bf 35}, C4-229

\bibitem{NKH} S. Nagato, P.H. Keeson and H.R. Harrison, 1979, Phys. Rev. {\bf B19}, 1633

\bibitem{TR} For other theoretical reviews the reader is referred to\\
 M. M\'ezard, G. Parisi and M.A. Virasoro, 1987 ``Spin Glass Theory and Beyond'', (World Scientific)\\
K.H. Fischer and J.A. Hertz, 1991 ``Spin Glasses'' (Cambridge)\\
D. Sherrington, 1990 in ``1989 Lectures on Complex Systems'' ed. E. Jen (Adison-Wesley), p415\\
D. Sherrington, 1992, in ``Electronic Phase Transitions'', ed. W. Hanke and Yu.V.Kopaev (North-Holland), p79

\bibitem{SS}D. Sherrington and B.W. Southern, 1975, J. Phys. {\bf F5}, 965

\bibitem{SK}D. Sherrington and S. Kirkpatrick, 1975, Phys. Rev. Lett {\bf 35}, 1972

\bibitem{S} D. Sherrington, 1980, J. Phys. {\bf A13}, 637

\bibitem{FH} R. Fisch and A.B. Harris, 1977, Phys. Rev. Lett. {\bf 37}, 756

\bibitem{BMI}B. Barbara, A.P. Malozemoff and Y. Imry, 1981, Phys. Rev. Lett {\bf 47}, 1852

\bibitem{P2} G. Parisi, 1980, J. Phys. {\bf A13}, L371

\bibitem{MB}P. Monod and H. Bouchiat, 1982, J. Phys. Paris) Lett. {\bf 43}, 145

\bibitem{MC} H. Maletta and P. Convert, 1979, Phys. Rev. Lett. {\bf 42}, 108

\bibitem{Suzuki} M. Suzuki, 1977, Prog. Theor. Phys. {\bf 58}, 1151

\bibitem{Chalupa} J. Chalupa, 1977, Solid State Comm. {\bf 22}, 315

\bibitem{AT}J.R. de Almeida and D.J. Thouless, 1978, J. Phys. {\bf A11}, 983

\bibitem{PR}E. Pytte and J. Rudnick, 1979, Phys. Rev. {\bf B19}, 3603

\bibitem{BM} A.J. Bray and M.A. Moore, J. Phys. {\bf C12}, 79

\bibitem{P} G. Parisi, 1979, Phys. Rev. Lett. {\bf 43}, 1754

\bibitem{RTV} R. Rammal, G. Toulouse and M.A. Virasoro, 1986, Revs. Mod. Phys. {\bf 58}, 765

\bibitem{MPSTV}M. M\'ezard, G. Parisi, N. Sourlas, G. Toulouse and M.A. Virasoro, 1984, J. Phys (Paris {\bf 45}, 843

\bibitem{BM2} A.J. Bray and M.A. Moore, 1984, J. Phys {\bf C17}, L463


\bibitem{KS}S. Kirkpatrick and D. Sherrington 1978, Phys. Rev. {\bf B17}, 4384

\bibitem{Y} A.P. Young, 1983, Phys. Rev. Lett {\bf 51}, 1206

\bibitem{BY} R.N. Bhatt and A.P. Young, 1985, Phys. Rev. Lett. {\bf 54}, 924\\
1988, Phys. Rev. {B37}, 5606

\bibitem{S2} D. Sherrington, 1986, Prog. Theor. Phys. Sup. 87, 180

\bibitem{S3} D. Sherrington, 1992, in ``Mathematical Studies of Neural Networks'', ed. J.G. Taylor (Elsevier), p261

\bibitem{CSH} A. Crisanti, H-J. Sommers and H. Horner, 1993, Z. Phys. {\bf B92}, 257

\bibitem{CK} L. Cugliandolo and J. Kurchan, 1993, Phys. Rev. Lett. {\bf 71}, 173

\bibitem{G} eg. W. Gotze, 1993, in ``Phase transitions and relaxation in Systems with Competing Energy Scales'', ed. T. Riste and D. Sherrington (Kluwer) p191

\bibitem{BCKM} J-P. Bouchaud, L.F. Cugliandolo, J. Kurchan and M. M\'ezard, 1998, in ``Spin glasses and random fields'' ed. A.P. Young (World Scientific) 

\bibitem{WS1} K.Y.M. Wong and D. Sherrington, 1990, J. Phys. {\bf A2}, 4659

\bibitem{van Kampen} N. van Kampen, 1992 ``Stochastic Processes in Physics and Chemistry'' (North Holland)

\bibitem{Parisi} G. Parisi, 1988 ``Statistical Field Theory'' (Addison-Wesley)

\bibitem{Peretto} P. Peretto, 1992 ``An Introduction to the Modelling of Neural Networks'' (Cambridge)

\bibitem{H} J.J. Hopfield, 1982, Proc. Natl. Acad. Sci. {\bf 79}, 2554

\bibitem{CS} A. Crisanti and H-J. Sommers, 1992, Z. Phys. {\bf B87}, 341

\bibitem{Struick} L.C.E. Struick, 1978, ``Physical Aging in Amorphous Polymers and Other Materials'' (Elsevier)

\bibitem{AGS} D.J. Amit, H. Gutfreund and H. Sompolinsky, 1985, Phys. Rev. Lett. {\bf 55}, 1530; Ann. Phys. {\bf 173}, 30

\bibitem{KGV} S. Kirkpatrick, C.D. Gelatt and M.P. Vecchi, 1983, Science {\bf 220}, 671

\bibitem{FA} Y.Fu and P.W. Anderson, 1986, J. Phys. {\bf A19}, 1605

\bibitem{KS2} I. Kanter and H. Sompolinsky, 1987, Phys. Rev. Lett. {\bf 58}, 164

\bibitem{WS} K.Y.M. Wong and D. Sherrington, 1987, J. Phys. {\bf A20}, L793

\bibitem{MP} M. M\'ezard and G. Parisi, 1986, J. Phys. (Paris) {\bf 47}, 1285

\bibitem{G1} E. Gardner, 1988, J. Phys. {\bf A21}, 257

\bibitem{GD} E. Gardner and B. Derrida, 1988, J. Phys. {\bf A21}, 271

\bibitem{WS2} K.Y.M. Wong and D. Sherrington, 1990, J. Phys. {\bf A23}, L175

\bibitem{KM} W. Krauth and M. M\'ezard, 1989, J. Phys. (Paris) {\bf 50}, 3057

\bibitem{CS1} A.C.C. Coolen and D. Sherrington, 1994, Phys. Rev. {\bf E49}, 1921

\bibitem{CS2} A.C.C. Coolen and D. Sherrington, 1994, J. Phys. {\bf A27}, 7687

\bibitem{SC} D. Sherrington and A.C.C. Coolen, 1995, in ``25 Years of
Non-Equilibrium Statistical Mechanics'' ed.~J.~J.~Brey et
al. (Springer)

\bibitem{LCS} S.N. Laughton, A.C.C. Coolen and D. Sherrington, 1996,
J. Phys. {\bf A29}, 763

\bibitem{TTCSS} M. Thomsen, M.F. Thorpe, T.C. Choy, D. Sherrington and H-J. Sommers, 1986, Phys. Rev. {\bf B33}, 1931

\bibitem{Young} A.P. Young, 1998, ``Spin Glasses and Random Fields'' (World Scientific)
 
\end{thebibliography}
\end{document}